%% file: paper_draft.tex
\documentclass[a4paper,fleqn,usenatbib]{mnras}
\usepackage{newtxtext}
\usepackage[T1]{fontenc}
\usepackage{ae,aecompl}
\usepackage{graphicx}
\usepackage{amsmath}
\usepackage{amssymb}
\usepackage{csquotes}
\usepackage{tablefootnote}
\usepackage{lipsum} 
\usepackage{soul}
\usepackage{xcolor}
\usepackage{multirow}

\newcommand{\mahcalib}{Navabi et al. ({\sl submitted.})}

\title[SMC Metallicity and Asymmetries]{Outside-In Evolution with a Twist: Metallicity Gradients and Asymmetries in the SMC}

\author[M. Navabi et al.]{
M. Navabi,$^{1}$\thanks{E-mail: m.navabi@surrey.ac.uk}
R. Carrera,$^{2}$
N. E. D. No\"el,$^{1}$
M. De Leo,$^{3,2}$
\\
$^{1}$Department of Physics, University of Surrey, Guildford GU2 7XH, Surrey, UK\\
$^{2}$INAF-Osservatorio di Astrofisica e Scienza dello Spazio, via P. Gobetti 93/3, 40129, Bologna, Italy\\
$^{3}$ Dipartimento di Fisica e Astronomia, Universita degli Studi di Bologna, Via Piero Gobetti 93/2, Bologna, 40129, Italy
}
\date{Accepted XXX. Received YYY; in original form ZZZ}

\pubyear{2025}

\begin{document}
\label{firstpage}
\pagerange{\pageref{firstpage}--\pageref{lastpage}}
\maketitle

\begin{abstract}
Taking advantage of the near-infrared calcium triplet lines, we determine metallicities for a sample of more than 3,500 red giant stars in the field of the Small Magellanic Cloud (SMC). We find a median metallicity of [Fe/H]=-1.05$\pm$0.01\,dex with a negative metallicity gradient of -0.064$\pm$0.007\,dex\,deg$^{-1}$ between 1\fdg2~to 6\fdg0~consistent with an outside-in evolution scenario. For the first time, we detect hints of a positive metallicity gradient within 1\fdg2, likely reflecting radial migration or centralised chemical enrichment. Azimuthal metallicity asymmetries are detected, with flatter gradients in the eastern and southern quadrants and steeper ones in the north and west. They are consistent with tidal interaction effects from the Large Magellanic Cloud (LMC). Finally, in spite of a clear distance and velocity bifurcations in the east, they seem to share a common chemical origin, in agreement with other studies.
\end{abstract}


\section{Introduction}
\label{sec:intro}

Situated at a distance of only $\sim$62.44$\pm$0.47\,kpc from us \citep[e.g.,][]{2020Graczyk}, the Small Magellanic Cloud (SMC) provides an unparalleled opportunity to study stellar populations in exquisite detail. As the less massive counterpart to the Large Magellanic Cloud \citep[LMC;][hereafter  Paper~I]{2002AJ....124.2639V,2024DeLeo}, the SMC has experienced substantial tidal disruption due to gravitational interactions with its larger neighbour \citep{Noel2013b,Noel2015,2017Carrera}. These interactions, along with the combined influence of the Milky Way \citep[MW;][]{Besla2010,2017MNRAS.468.3428P,2021Tatton}, have shaped the SMC’s structure, triggering the formation of the Magellanic Bridge, Stream, and Leading Arm \citep[e.g.,][]{1989MNRAS.241..667H,Olsen2011,Noel2013b,Noel2015,2017Carrera,2003ApJ...586..170P,2008Nidever, 2010Nidever,2013Nidever,2011PASA...28..117D, 2012MNRAS.421.2109B, 2016ARA&A..54..363D, 2019MNRAS.482L...9B,2021Youssoufi,2021Tatton}. These dynamical encounters have also profoundly influenced the SMC’s star formation history, which exhibits an episodic, burst-like pattern \citep[e.g.,][]{Noel2009,2024Sakowska}. Such bursts imprint distinct chemical signatures on the galaxy’s stellar populations, offering critical insights into its evolutionary trajectory.

Despite extensive study, the SMC’s chemical evolution remains incompletely understood. Investigations using stellar clusters \citep[e.g.,][]{Piatti2007,Parisi2009} reveal a complex enrichment history: an early rise in metallicity, followed by non-monotonic variations, including a possible dip between 10 and 3\,Gyr ago and a renewed increase in the last 2–3\,Gyr \citep[e.g.,][]{Parisi2015,Parisi2022,Bortoli2022}. However, discrepancies persist between cluster and field star populations. While clusters exhibit a bimodal metallicity distribution \citep[e.g.,][]{Parisi2015,Bortoli2022}, field stars do not \citep[e.g.,][]{Carrera2008,2018Choudhury}. Additionally, field stars display a radial metallicity gradient \citep[e.g., -0.075$\pm$0.011\,dex\,deg$^{-1}$ within 4\fdg0][hereafter D14]{2014dobbie}, whereas clusters show a flat trend \citep[e.g.,][]{Piatti2007,Parisi2015}. Recent work suggests these gradients may vary regionally, likely due to LMC interactions \citep[e.g.,][]{LI2024}.

A comprehensive understanding of the SMC’s chemical evolution requires detailed abundance studies across multiple elements, each tracing different nucleosynthetic pathways and timescales. However, high-resolution spectroscopic analyses remain challenging at the SMC’s distance, limiting most studies to clusters \citep[e.g.,][]{mucciarelli2023c} or small field-star samples \citep[e.g.,][]{vanderswaelmen2013}, including those from the Apache Point Observatory Galactic Evolution Experiment \citep[APOGEE;][]{2017Majewski,nidever2020}.

The infrared \ion{Ca}{ii} triplet (CaT, around 850\,nm) offers a powerful alternative, enabling metallicity measurements for large stellar samples \citep[e.g.,][]{Carrera2007}. Leveraging this method, we present a homogeneous analysis of SMC stars spanning the entire galaxy in order to improve our knowledge of the chemical evolution of this galaxy linking it with the interactions with its larger companion. The paper is structured as follows. In Sect.~\ref{sec:obs}, we describe the data sets used in our analysis.
In Sect.~\ref{sec:analysis}, we outline our process for deriving stellar metallicities. We present the measured metallicity distribution function of the SMC and examine regional variations, particularly focusing on potential radial and azimuthal gradients of the SMC in Sect.~\ref{sec:res}. These findings are discussed in conjunction with distance and kinematic trends in Sect.~\ref{sec:disc}, offering insights into the effects of SMC-LMC interactions. Additionally, we present a map of the metal-poor stellar population in the SMC and compare its kinematic distribution with that of the overall stellar population.
Finally, in Sect.~\ref{sec:conclusions}, we summarize our key results and conclusions.

\begin{figure*}
	\includegraphics[width=\textwidth]
    {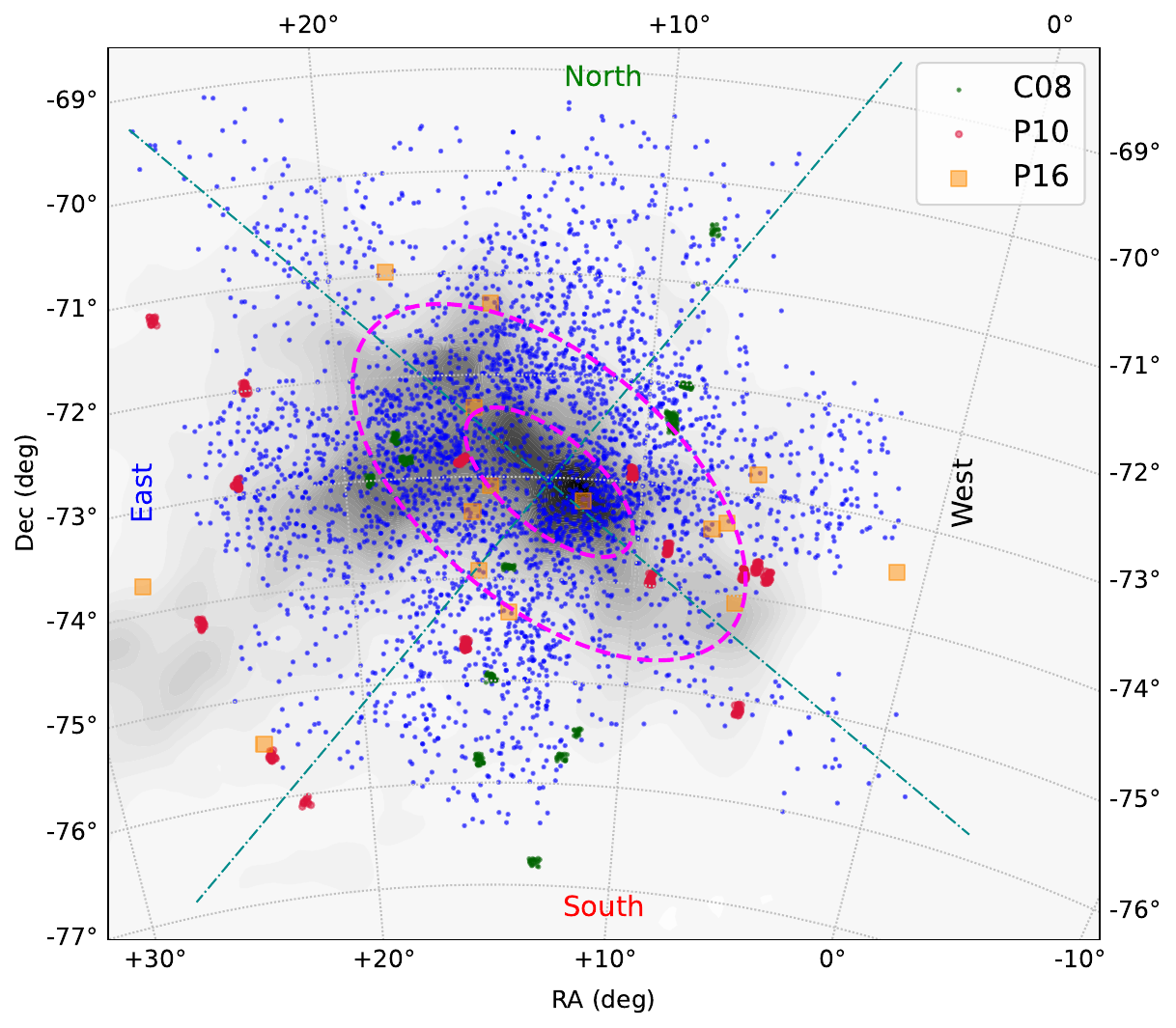} 
    \caption{Spatial distribution of our sample (blue points) overimposed to the \ion{H}{i} map \citep{2015Kalberla}. The location of the objects studied by C08: \citet[][in green]{Carrera2008} and P10 and P16: \citet[][in red and yellow, respectively]{Parisi2010, Parisi2016}. The four quadrants and the two example ellipses used in our analysis are also shown as references (see text for details).}
    \label{fig:map}
\end{figure*}

\section{Observational material}
\label{sec:obs}
The starting dataset for the present analysis is the sample studied in Paper~I in which we obtained medium resolution spectra, R$\sim$8,500, in the CaT region, $\sim$850\,nm for 2,680 stars in the line of sight of the SMC. These spectra were acquired using the grating 1700D in the red arm of the AAOmega spectrograph \citep{2004SPIE.5492..389S} fibre-fed by the Two Degree Field (2dF) multi-object system installed at the Anglo-Australian 4\,m telescope (AAT, Siding Spring Observatory, Australia). The target stars were selected from the expected position of the upper red giant branch in the colour-magnitude diagrams from the Two Micron All Sky Survey Two Micron All Sky Survey \citep[2MASS;][]{2006AJ....131.1163S}. The raw data were processed through the 2dF data reduction pipeline \citep[2dfdr2;][]{Sharp2010}, which performs the initial data reduction steps such as bias subtraction, flat-field normalization, fibre tracing and extraction, and wavelength calibration. The sky subtraction and stacking were performed following the procedure described by \citet{carrera2017}. Finally, we considered the stars' radial velocities obtained in Paper~I (see their Table~A1).
Our sample was complemented with that published by D14, who obtained spectra for 4,197 stars in the SMC line of sight using the same instrumental configuration and similar selection function as Paper~I. For the present work, we downloaded the raw data directly from the AAT data archive\footnote{\url{https://archives.datacentral.org.au/}}. The spectra were processed following the same procedure conducted in Paper~I but considering the radial velocities obtained by D14. A detailed comparison between both sets of radial velocities can be found in Paper~I.

All together, our extended sample contains spectra for  6,877 stars. We removed low-quality spectra with signal-to-noise ratio (S/N) lower than 15\,pix$^{-1}$ and conducted a visual inspection, excluding stars which are likely to be Asymptotic Giant Branch (AGB) stars. In the case of the D14 sample, we took into account the classification provided by these authors. We then discarded foreground contaminants, selecting only stars with radial velocities in the range 65 $\leq v_{\rm rad}$ [\,km\,s$^{-1}$] $\leq$ 230. The final sample contains 3,697 spectra (1,291 from Paper~I and 2,332 from D14) with fourteen stars targeted by both datasets, which will be used for homogeneity checks (see Sect.~\ref{sec:analysis}).  Figure~\ref{fig:map} illustrates the spatial distribution of our final sample.

In this work, we also take advantage of the \textit{Gaia} EDR3 catalogue \citet{Gaia2021}. To cross-match our sample with the \textit{Gaia} one, we use two different methods. The first approach employs \textsc{Astropy} \citep{2013Astropy,2018Astropy} for cross-matching both catalogues within an initial search radius of one arcsecond, progressively increasing the radius to two and three arcseconds to identify potential additional matches. However, no additional matches were identified beyond the initial radius. There are 6,843 stars in our catalogue found
in \textit{Gaia} one.
The second method utilizes the \textit{Gaia} Archive's cross-matching functionality \footnote{https://gea.esac.esa.int/archive/} based on 2MASS identifiers. Although this method could not find matches for all stars in our sample, it served as a verification tool for the results obtained through \textsc{Astropy}. Of the 6,877 stars in our initial dataset, the \textit{Gaia} Archive identified 4,234 stars with corresponding 2MASS IDs. A comparison of the 2MASS IDs from the two methods revealed that they are consistent for all but 79 stars. Therefore, we use the first method to have the \textit{Gaia} parameters for most stars.

\unskip
\input{Tables/Table_validation}
\unskip

\begin{figure}
    \includegraphics[width=\columnwidth]{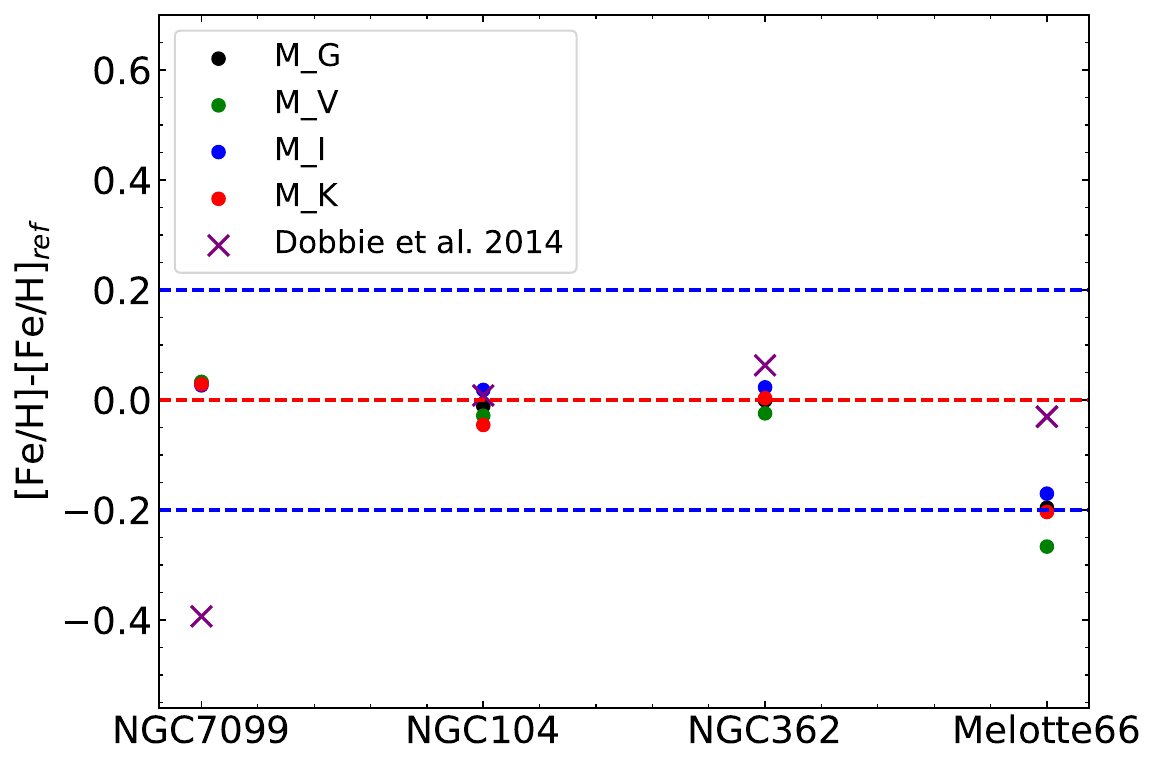}
    \caption{Residual metallicities of each cluster across the four bands. The cross symbols illustrate the discrepancies between the metallicities in this work and those of D14, both using the $K_s$ band. The blue dashed lines are the standard deviation of the differences of the CaT calibration (Fig.~3 of \mahcalib.}
    \label{fig:robustness_clusters}
\end{figure}

\section{Metallicity determination}
\label{sec:analysis}

We determined the metallicity of stars in our sample from the strength of CaT lines using the revised calibration obtained by \mahcalib~which relates both quantities taking into account also the luminosity of the stars. The procedure we followed is fully described by \citet{Carrera2007}, \citet{Carrera2013}, and \mahcalib. The strength of each line is determined from its profile, which is characterised by a combination of a Gaussian and a Lorentzian functions within a given band-pass \citep{1997Rutledgeb,1997Rutledgea,Cole2004}. The continuum is also measured in band-passes specifically selected for this purpose.

\mahcalib~obtained their calibration for several luminosity indicators, of which we use here the absolute magnitude in the 2MASS $K_s$ and $G$ bands. For this purpose, we de-redden the apparent magnitudes using the values $E(B-V)$ and their respective uncertainties for each star provided by the \textsc{dustmaps} package \citet{Planck2016}. These values are converted into extinctions in both bands, $A_{\lambda}$, using an extinction coefficient of 0.114\,mag \citep{1989ApJ...345..245C}. We assume a distance modulus for the SMC of $(m-M)_0$ = 18.86$\pm$0.11\,mag \citep{2021Martinez}.

Following the prescriptions from \mahcalib~
and \citet{Starkenburg2010}, the metallicity of stars as a function of the strengths of the CaT lines and the absolute magnitude in each $\lambda$ band, $M_{\lambda}$, is given by:
\[
    [Fe/H] = a + b \times M_{\lambda} +c \times \Sigma Ca +d \times \Sigma Ca ^{-1.5} + e \times M_{\lambda} \Sigma Ca
    \label{eq:metal}
\]
where $\Sigma Ca$ is the sum of the strengths of the three CaT lines for each star. The coefficients \textit{a}, \textit{b}, \textit{c}, \textit{d}, and \textit{e} have been computed
using the reference metallicities listed in Table~1 from \mahcalib.
The metallicity for each star is determined from the above equation using a Monte Carlo approach, which also provides a realistic estimation of its uncertainty. In short, we consider each input variable as a Gaussian probability distribution centred on its nominal values, with a standard deviation corresponding to its uncertainties. The considered variables are: $\Sigma Ca$, M$_{\lambda}$, reddening, and distance modulus. The value for each of them is selected randomly within this probability distribution in order to derive the corresponding metallicity. This procedure is repeated 10,000 times, and the resulting metallicities for each star are distributed following a Gaussian function again. The mean of this distribution for a given star is the derived metallicity, while the width of the distribution provides an estimation of its uncertainty.

There are fourteen common stars between the initial sample from Paper~I and D14. To validate our results, we calculated the metallicity of these stars before removing one of the duplicate entries from our catalogue. We follow the same procedure for the 86 stars targeted twice within Paper~I's fields. In both cases, the histograms of the metallicity differences follows a Gaussian distribution, peaking near zero with a standard deviation of approximately 0.3\,dex. For each pair of duplicates, the star with superior spectral quality or higher signal-to-noise ratio is retained in the final sample.

To check the reliability of the metallicity determination presented here, we study four stellar clusters, which cover a broad range in metallicities, observed by D14 using the same instrumental configuration as the SMC targets. We reduced these stars from scratch using the same procedure as the case of the SMC data, applying similar quality constraints to the SMC spectra. For each system, the members are selected from their radial velocities within $\pm$10\,km\,s$^{-1}$ from the mean value provided by D14 for each of them.
Once more, the metallicity of each cluster star has been derived following the same procedure as for the SMC stars: the metallicity is the median of the values obtained for all the members of each cluster after applying a 3-$\sigma$ clipping rejection. The obtained values, together with their respective uncertainties, are listed in Table~\ref{tab:fe}. This is important to note that, within the uncertainties, the values are in good agreement.

Figure~\ref{fig:robustness_clusters} presents the residual metallicity values (i.e., the differences between the high-resolution reference and our results) for the globular clusters across four magnitudes: $M_I$, $M_K$, $M_V$, and $M_G$. These residuals generally fall within the typical uncertainties of the CaT method obtained by Fig.3 in \mahcalib, indicating good agreement. Melotte~66, however, stands out due to its high-resolution reference value being derived solely from \citep{2014Carraro}, as it has not been included in surveys like OCCASO \citep[Open Clusters Chemical Abundances from Spanish Observatories; ][]{2024Carbajo}, APOGEE \citep{2022Myers}, or GES \citep[Gaia-ESO Survey; ][]{2022A&A...666A.121R}. Consequently, this cluster shows the largest discrepancy, exceeding $\sim 0.2$\,dex in the $V$ band. However, the $K_s$ band is particularly significant for this study and demonstrates the best agreement with the high-resolution reference values.
Also, D14 derived the CaT metallicity in the $K_s$ band for these clusters M30, NGC~362, NGC~104, and Melotte~66, with reported values of -1.91$\pm$0.07, -1.13$\pm$0.08, -0.75$\pm$0.07, and -0.47$\pm$0.09\,dex, respectively. The cross symbols in Fig.~\ref{fig:robustness_clusters} represent the discrepancies between our results and the D14 values in the $K_s$ band. Those discrepancies are less than $0.06$\,dex for NGC~104, NGC~362, and Melotte~66, while for NGC~7099 is about $0.43$\,dex. Nonetheless, our results show minimal deviation from the high-resolution spectroscopy determinations for the latter cluster. 

\begin{figure}
    \centering
\includegraphics[width=\linewidth]{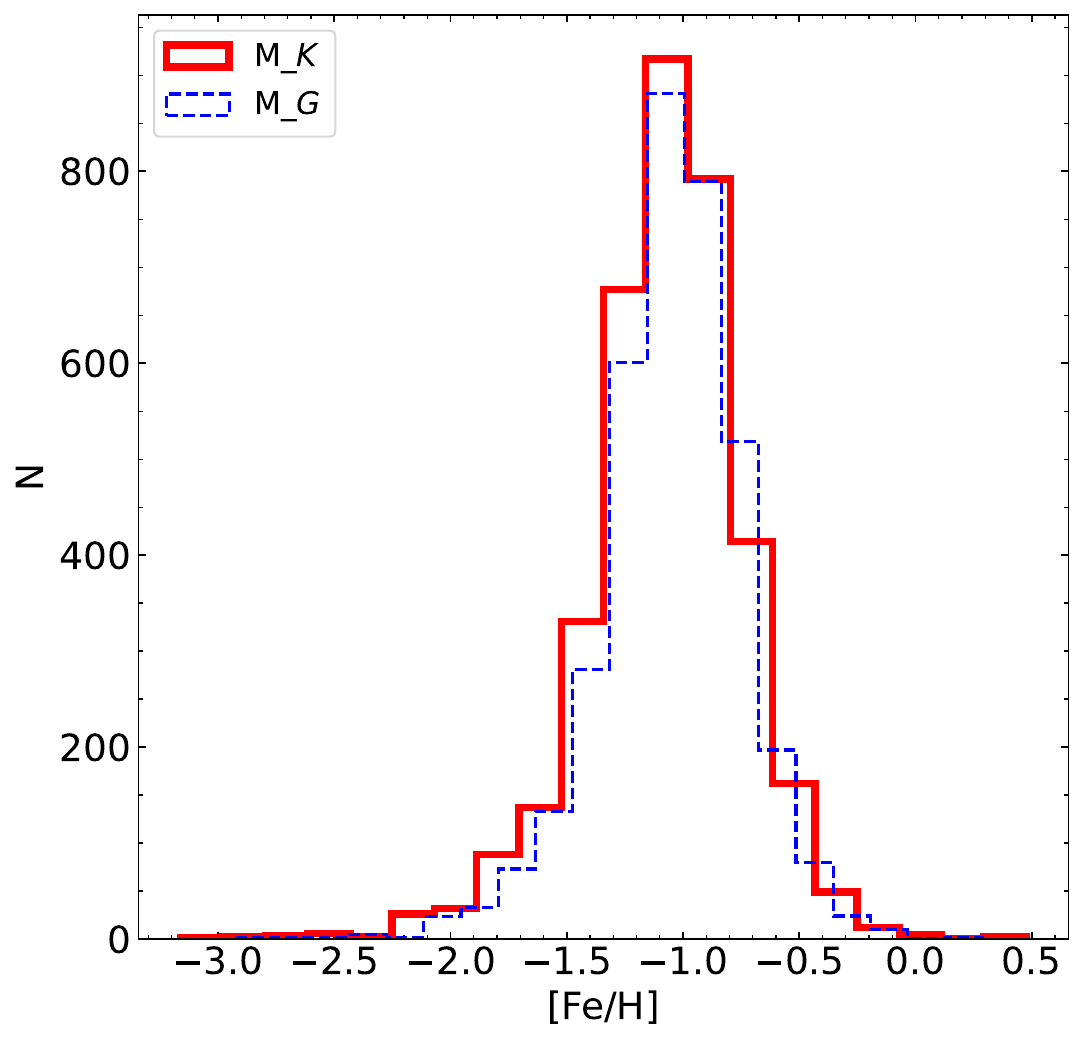}
    \caption{The metallicity distribution of our sample with both \textit{K$_s$} (red solid line) and \textit{Gaia} \textit{G} (blue dashed line) magnitudes. The details of these distributors are summarized in Table~\ref{tab:metallicity_distributions}.}
    \label{fig:hist_metal}
\end{figure}
\input{Tables/metal_distribution}
\input{Tables/metal_literature}

\begin{figure*}
    \centering
    \includegraphics[width=\linewidth]{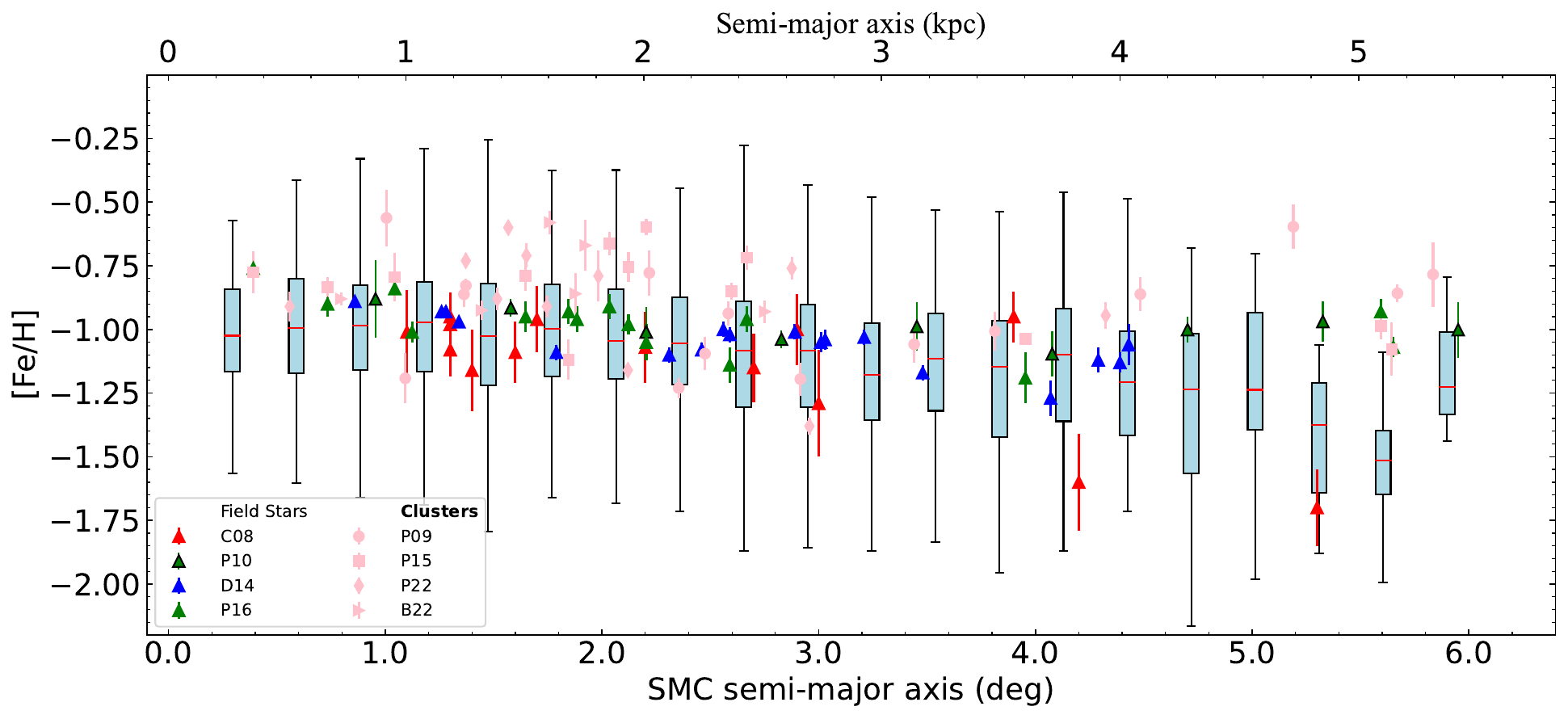}
    \caption{The metallicity gradient across semi-major axis of the ellipses as described in Sec.\ref{sec:radial}. For comparison, we include the median metallicities of field stars from C08, P10, P16: \citet[]{Carrera2008}, \citet[]{Parisi2010}, \citet[ ]{Parisi2016}, D14 and clusters from P09, P15, P22, B22: \citet[]{Parisi2009},  \citet[]{Parisi2015}, \citet[]{Parisi2022}, and \citet[]{Bortoli2022}, respectively (see inset). The top of the x-axis shows the corresponding distance in kpc.}
    \label{fig:gradient_metal_proj}
\end{figure*}

\section{The metallicity content of the SMC}
\label{sec:res}

Figure~\ref{fig:hist_metal} presents the derived metallicity distributions for our sample, computed using \textit{K$_s$} (red solid histogram) and \textit{Gaia G} (blue dashed histogram) magnitudes.
Both distributions are unimodal and exhibit a high degree of similarity. A Kolmogorov–Smirnov (KS) test yielded a statistic of 0.6 and a \textit{p}-value of zero, indicating no significant differences between the two distributions and confirming the consistency of our metallicity determinations.
Their medians, 10th (P10) and 90th (P90) percentiles, and the standard deviations are listed in the first two rows of Table~\ref{tab:metallicity_distributions}. Since the metallicities derived from both \textit{K$_s$} and \textit{Gaia G}-band magnitudes are very similar, we base our analysis solely on the values obtained from the \textit{K$_s$} magnitudes, due to the greater number of stars in this band. However, we have verified that using the \textit{G}-band yields nearly identical results.

\subsection{Radial Metallicity Gradient}
\label{sec:radial}

To determine whether a radial metallicity gradient exists in our sample and further investigate the spatial distribution of metallicities, we first de-project the positions of the stars in our sample following \citet{2009Cioni}.
In short, we first calculate the position angle ($\phi$) and the angular distance ($\rho$) in the sky for each star from the SMC centre, and subsequently convert these values into Cartesian coordinates using the following equations: 
\[
 \begin{split}
x(\alpha,\delta) = \rho~\sin{\phi} \\
y(\alpha,\delta) = \rho~\cos{\phi} \\ 
 \end{split}
\]

We then rotate the stars' coordinates with respect to the position of the SMC's line of nodes that corresponds to PA$_{SMC}$ = 130\degr~\citep{2012Subramanian}:

\[
\begin{split}
x^{\prime} = x~\cos{PA_{SMC}} + y~ \sin{PA_{SMC}}\\
y^{\prime}= -x~\sin{PA_{SMC}} + y~ \cos{PA_{SMC}}
\end{split}
\]

To calculate the projected distance, we utilise the equation:
\[
r= \sqrt{x^{\prime 2} + y^{\prime 2}}
\]
Note that we only consider the projected coordinates of $x^{\prime}$ and $y^{\prime}$ and neglect the depth of the puzzling line of sight of the SMC \citep[see, e.g.,][]{2024Sakowska}.

To obtain the SMC's radial metallicity distribution we overlay twenty concentric ellipses with an axial ratio of $a/b$=1.5 \citep{2012Subramanian} centred on the red clump centre of the SMC \citep{2012Subramanian} with coordinates  (RA, Dec) $=$ (13\fdg143, -73\fdg05), and the semi-major axis, $a$, increasing incrementally by 0\fdg3.
Two such ellipses are shown in Fig.~\ref{fig:map} as a reference. 
Within each elliptical annulus, we apply a bootstrap sampling method to estimate the median metallicity and its associated uncertainty. The results of this procedure are presented in  Table~\ref{tab:radial}, which also lists the 10th and 90th percentiles for each of them. The last column is the standard deviation of the [Fe/H] distribution in each bin.

Figure~\ref{fig:gradient_metal_proj} illustrates the variation of the metal content of our field stars' sample as a function of the semi-major axis. For each annulus, the box (blue rectangles) goes from the 1st to the 3rd quartiles with the red line representing the median (see Table~\ref{tab:radial}).
The whiskers extend to the minimum and maximum values, excluding outliers. 
The figure reveals a distinct break in the SMC's metallicity gradient. The profile is nearly flat or slightly positive from the centre out to a radius of approximately 1$\fdg$2, beyond which it transitions to a clear negative gradient. This pattern is mirrored by the 1st and 3rd quartile values, underscoring the feature's robustness. To test the statistical significance of this break, we performed a piecewise linear fit, treating the breakpoint as a free parameter. An MCMC analysis robustly locates this break at 1$\fdg$2$\pm$0$\fdg$3. The slopes for the inner and outer regions are listed in  Table~\ref{tab:mg}. An F-test comparing this broken model to a single linear fit yields $F$=2.9 ($p$=0.02), confirming that the two-segment model is a statistically better description of the data. This result provides strong evidence for a genuine transition in the SMC's metallicity distribution at this radius.

To statistically validate the significant positive metallicity gradient within the central 1\fdg2, we applied a bootstrap sampling. In brief, from all the stars inside 1\fdg2, we randomly select half of them. The objects in this subsample are distributed in the same elliptical annuli as before according to their distances, determined by the median, standard deviation, etc. of the derived metallicity distributions for each of them. Finally, we perform a weighted linear fit in order to estimate the slope of the gradient. This procedure is repeated 500 times, and the median of the derived slopes suggests, at a two $\sigma$ level, the existence of a positive gradient in the central part of this galaxy. This finding agrees with the MCMC result within the uncertainties.

Outside 1\fdg2, the metallicity clearly decreases as we move away from the centre in the innermost $\sim$3\fdg5, which several fluctuations from there off probably related to azimuthal variations (see Sect.~\ref{sec:azimuth}). In order to investigate the nature of this trend, we parametrise it with two linear functions, considering also the transition point. To do that we implement chi-square residual minimization using \textsc{lmfit} package \citep{lmfit}.
We find that inside 5\fdg0$\pm$0\fdg2 there is a shallow decrease of metalliticity with a rate of -0.06$\pm$0.01\,dex\,deg$^{-1}$ (-0.06$\pm$0.01\,dex\,kpc$^{-1}$), while outside this value the gradient is more steeper, with a slope of -0.4$\pm$0.7\,dex\,deg$^{-1}$, although this value is based only on three median points. The outermost distance analysed, at 6\fdg0, exhibits a slight increase in metallicity, but this result is based only on three stars.

\unskip
\input{Tables/radial}
\unskip

\begin{figure}
     \centering
        \includegraphics[width=\linewidth]{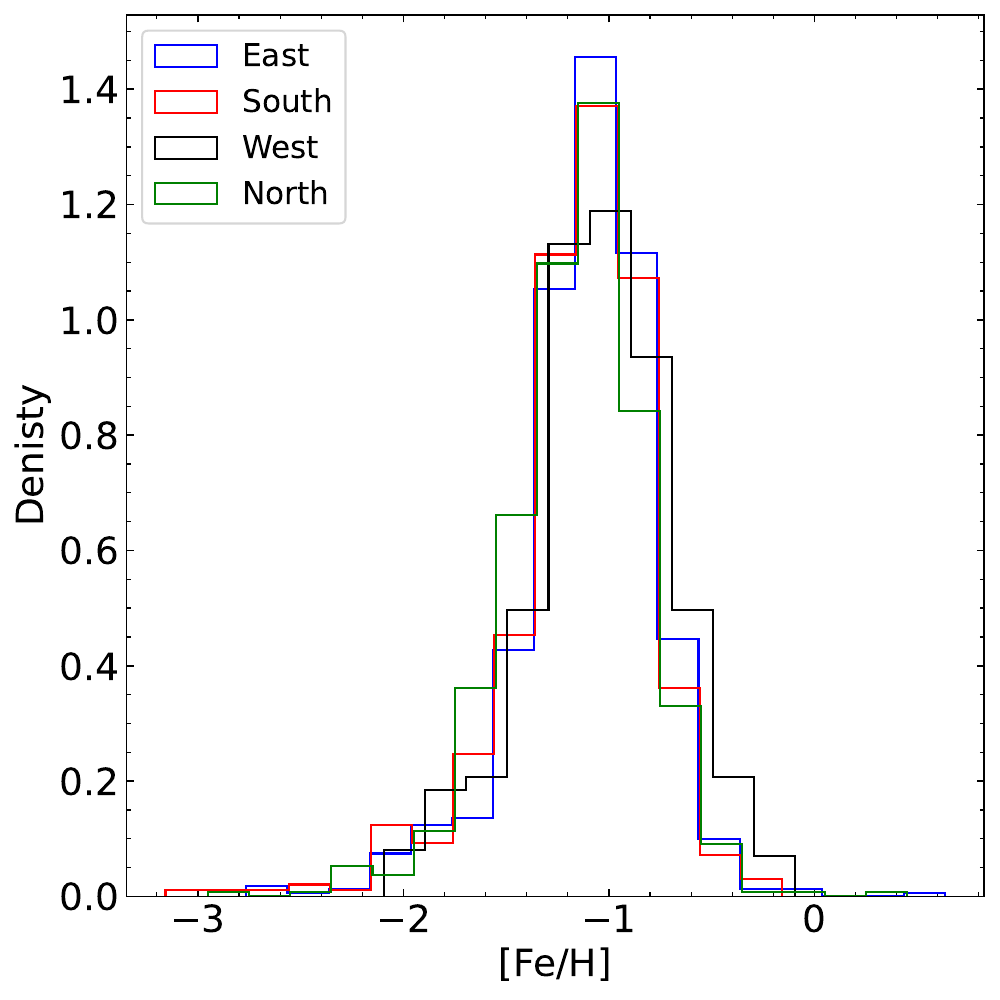}
    \caption{Normalised [Fe/H] distributions for RGB stars in the four quadrants of the SMC: East (blue), South (red), West (black), and North (green).
    A bootstrap sampling method was applied here to determine the statistics for each quadrant listed in Table~\ref{tab:metallicity_distributions}. }
    \label{fig:azimuth_hist}
\end{figure}

\begin{figure*}
     \centering
    \includegraphics[width=\linewidth]{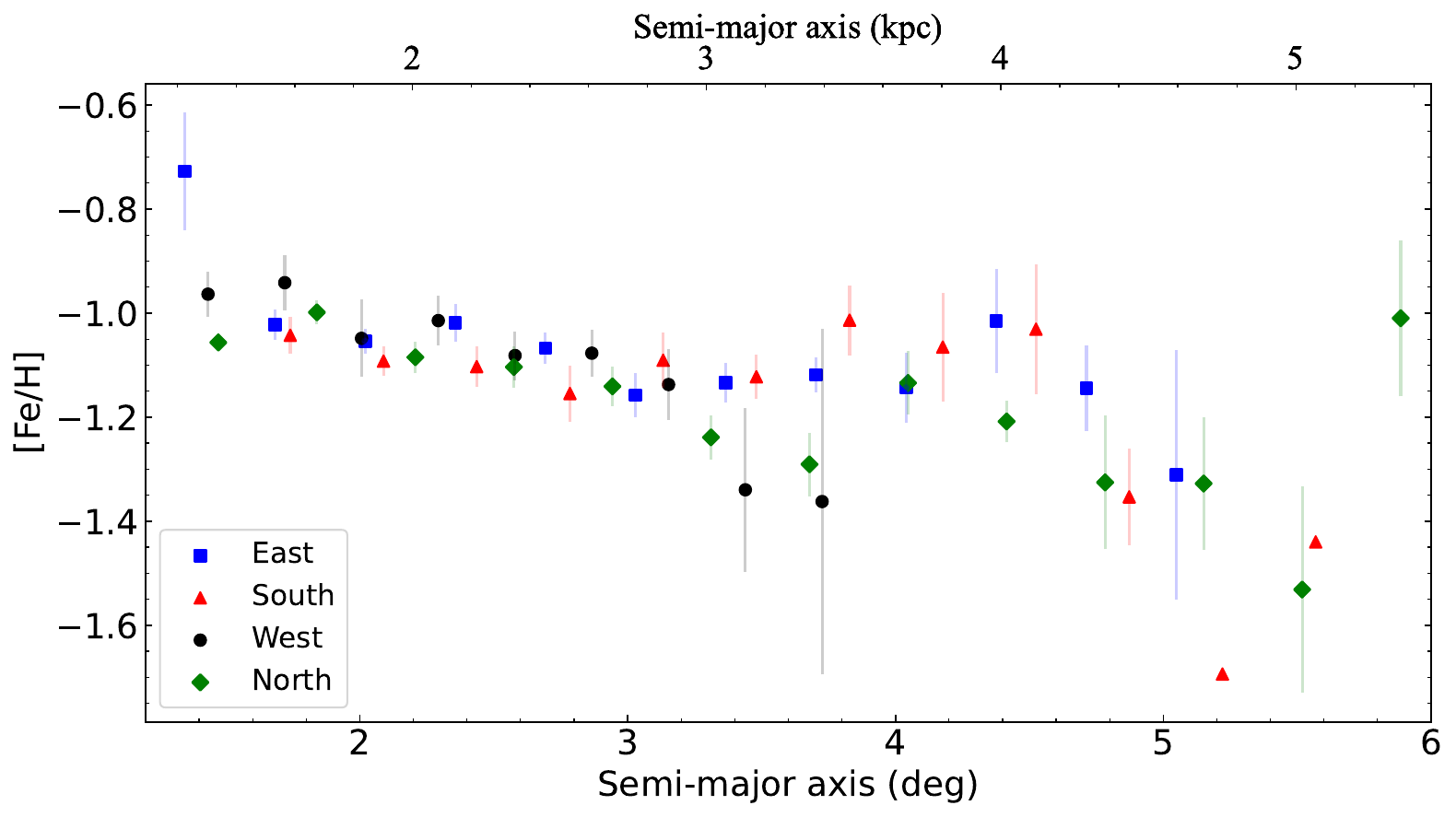}
             \caption{
    Radial metallicity distribution within each quadrant following the methodology described in Sect.~\ref{sec:radial}. The inner 1\fdg2 region is omitted due to low star counts in some quadrants; gradients within this region are analysed globally in Section 4. Table~\ref{tab:azimuth} lists the bin size and the number of stars contributing to each radial bin in each quadrant.}

    \label{fig:azimuth_radial}
\end{figure*}

\unskip
\input{Tables/MG}
\unskip

\subsection{Azimuthal Metallicity variations}
\label{sec:azimuth}

As commented above, the metallicity distribution shows some fluctuations between 3\fdg5 and 6\fdg0. In order to investigate whether they are related to azimuthal variations of metallicity, which is possible due to the spatial coverage and size of our sample,
we divide the SMC into quadrants as shown in Fig.~\ref{fig:map}: North, East, West, and South. The central region, inside 1\fdg2, is excluded from this analysis to minimize potential biases caused by the strong central concentration of stars, as well as the outskirts (> 3\fdg5), where more significant differences can be observed.
The normalized metallicity distributions for all quadrants are presented in Fig.~\ref{fig:azimuth_hist} 
We apply a bootstrap sampling method to determine the statistics for each quadrant listed in Table~\ref{tab:metallicity_distributions}. 
The northern quadrant, in green, exhibits a distinct excess of stars with metallicities between -1.5 and -2\,dex and a smaller 10th percentile (see Sect.~\ref{sec:mpm} for a further discussion). In contrast, the western region has an excess of metal-rich stars, -0.5$\leq$[Fe/H]$\leq$0\,dex, and a higher 90th percentile by at least 0.15\,dex.

Figure~\ref{fig:azimuth_radial} illustrates the radial metallicity distribution within each quadrant following the methodology described in Sect.~\ref{sec:radial}. Table~\ref{tab:azimuth} shows the number of stars in each radial bin for every quadrant.
A clear distinction is observed between the inner (<3\fdg8) and outer (>3\fdg8) regions relative to the SMC tidal radius, which has been reported as 3\fdg4 and 4\fdg5 by \citet{2021Dias,2020Massana}, respectively.
In the inner region, the radial metallicity gradient exhibits a nearly identical pattern in the four azimuthal regions up to 3\fdg2. Between 1\fdg2 and 3\fdg8, the gradient is steeper in the western and northern quadrants compared to the overall trend, while the eastern and southern quadrants show a more gradual decline (see Table~\ref{tab:mg}).
Beyond 3\fdg8, the eastern, northern, and southern regions display a similar pattern with a peak before undergoing a sudden decline up to 5\fdg6. In contrast, the western region, located opposite the LMC, spans a smaller area, extending only to 3\fdg8, a trend consistent with the behaviour of western halo clusters \citep[e.g.][]{Parisi2022}.

\unskip
\input{Tables/azimuth}

\unskip

\section{Discussion}
\label{sec:disc}

\subsection{Outside-In Formation with Central Inversion?}

Our measurements confirm a negative radial metallicity gradient in the SMC between 1$\fdg$2 and 6\fdg0, consistent with previous findings. The median gradient of $-0.064\pm0.007$\,dex\,deg$^{-1}$ aligns with expectations for dwarf irregular galaxies undergoing outside-in formation \citep[e.g.,][]{apariciotikhonov2000,Gallart2008,McQuinn2017}. The outermost regions (>5\fdg0) lack stars with [Fe/H]~$> -0.6$\,dex, implying an older stellar population formed $\sim12$\,Gyr ago \citep{Carrera2008}, whereas intermediate-age and younger stars with [Fe/H]~$\sim -0.6$ to $-0.5$\,dex are more centrally concentrated, associated with a star formation burst around $\sim$1\,Gyr \citep{mucciarelli2023}.
This gradient and the progressive central concentration of more metal-rich populations reflect a broader outside-in chemical evolution scenario, consistent with studies of other irregular systems such as WLM, IC~1613, Aquarius, NGC~6822, and Leo~A \citep{2013Leaman,2013Kirby,2017Kirby,2022Taibi}.

However, perhaps the most remarkable result is that, within the central $\leq$1$\fdg$2, we detect a positive metallicity gradient—the first such evidence reported using field stars in the SMC. This central inversion contrasts with the expected outside-in trend and resembles phenomena observed in the inner regions of the LMC and M33 \citep{2007Barker,2011Barker,2009Williams}. These inner metallicity inversions are often attributed to stellar radial migration, a process supported by simulations of both LMC- and MW–mass disks \citep{2008Ro,2012Ro,2012Radburn}. 
The statistical significance of this positive central gradient is marginal, and the result should be interpreted as tentative. Additional spectroscopic data in the inner regions of the SMC will be needed to confirm or refute this trend and to properly quantify its impact on the SMC's chemical structure.

We also examined the two-dimensional metallicity distribution across four azimuthal sectors. The inner 1$\fdg$2 region is excluded due to low star counts in certain quadrants.
This is illustrated in Fig.~\ref{fig:azimuth_hist} and~\ref{fig:azimuth_radial}, which show the median metallicity profiles and corresponding gradients in each quadrant.
Although steeper radial gradients are observed in the western and northern quadrants out to $\sim3\fdg8$-potentially due to tidal effects from the LMC on the eastern side of the SMC \citep{2013Nidever,2020Choudhury,2020deleo}-this finding is not statistically significant. A Kolmogorov-Smirnov test reveals no significant difference compared to the other quadrants at the 95\% confidence level.

Our findings reinforce earlier cluster-based results showing a consistent western gradient with no inversion \citep{Parisi2022}.
Taken together, these results reveal a complex evolutionary narrative: the SMC exhibits a dominant outside-in formation pattern, overprinted by central migration-driven inversion and external tidal disturbances that have shaped its metallicity structure in both radial and azimuthal dimensions. This two-dimensional chemical map emphasizes the need to interpret metallicity gradients not as simple linear trends but as dynamic, interaction-modulated signatures of galactic evolution.

A key contribution of our study is the detailed examination of azimuthal variations in the metallicity gradient. By dividing the SMC into quadrants, we identify a significant asymmetry: the western regions show a pronounced metallicity decline with radius, whereas the eastern side—facing the Magellanic Bridge—maintains a nearly flat profile out to $\sim$4\fdg0. This finding spectroscopically confirms the radial asymmetry reported by \citet{2020Choudhury}, and points to population mixing caused by tidal interactions with the LMC. The removal or redistribution of metal-rich stars from the eastern SMC, likely through tidal stripping \citep{2013Nidever,2020deleo,2021Grady, Almeida2024}, may have diluted any underlying gradient. In contrast, the western SMC—less affected by such interactions—retains a steeper and uninterrupted gradient \citep{Parisi2022,2022Dias}. This azimuthally resolved perspective provides a richer, more complete understanding of the SMC’s evolution, reinforcing the importance of combining radial and angular diagnostics in galactic archaeology.

\begin{figure*}
    \centering
    \includegraphics[width=\textwidth]{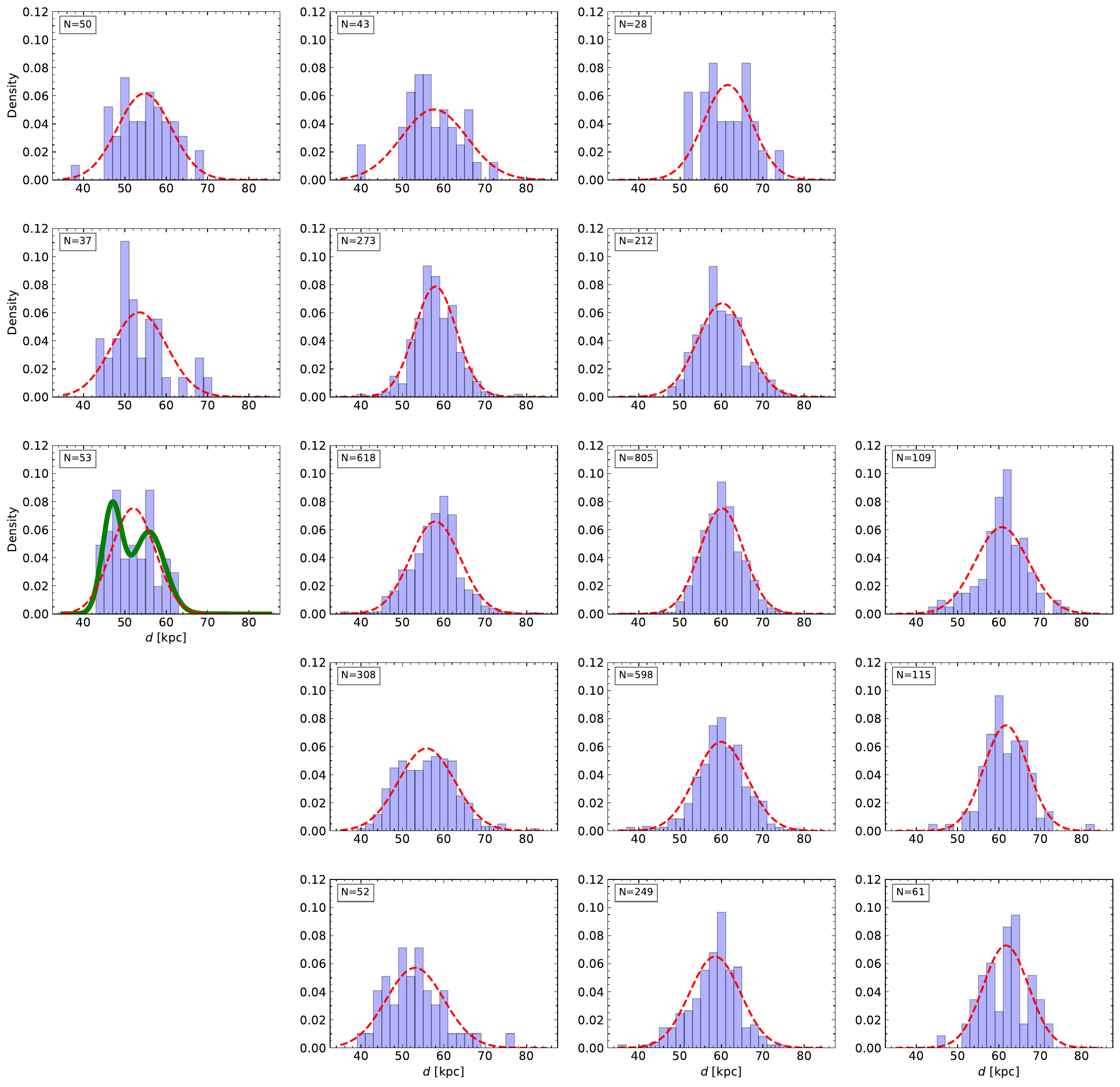}
    \caption{Distance histograms of the HEALPix subsamples of SMC RGB stars, with red dashed lines showing Gaussian functions based on the mean and standard deviation of each distribution. The Eastern subsample displays a double-Gaussian profile, indicated by green lines.}   
    \label{fig:fields_dis}
    \end{figure*}
\begin{figure*}
    \centering
     \includegraphics[width=\textwidth]{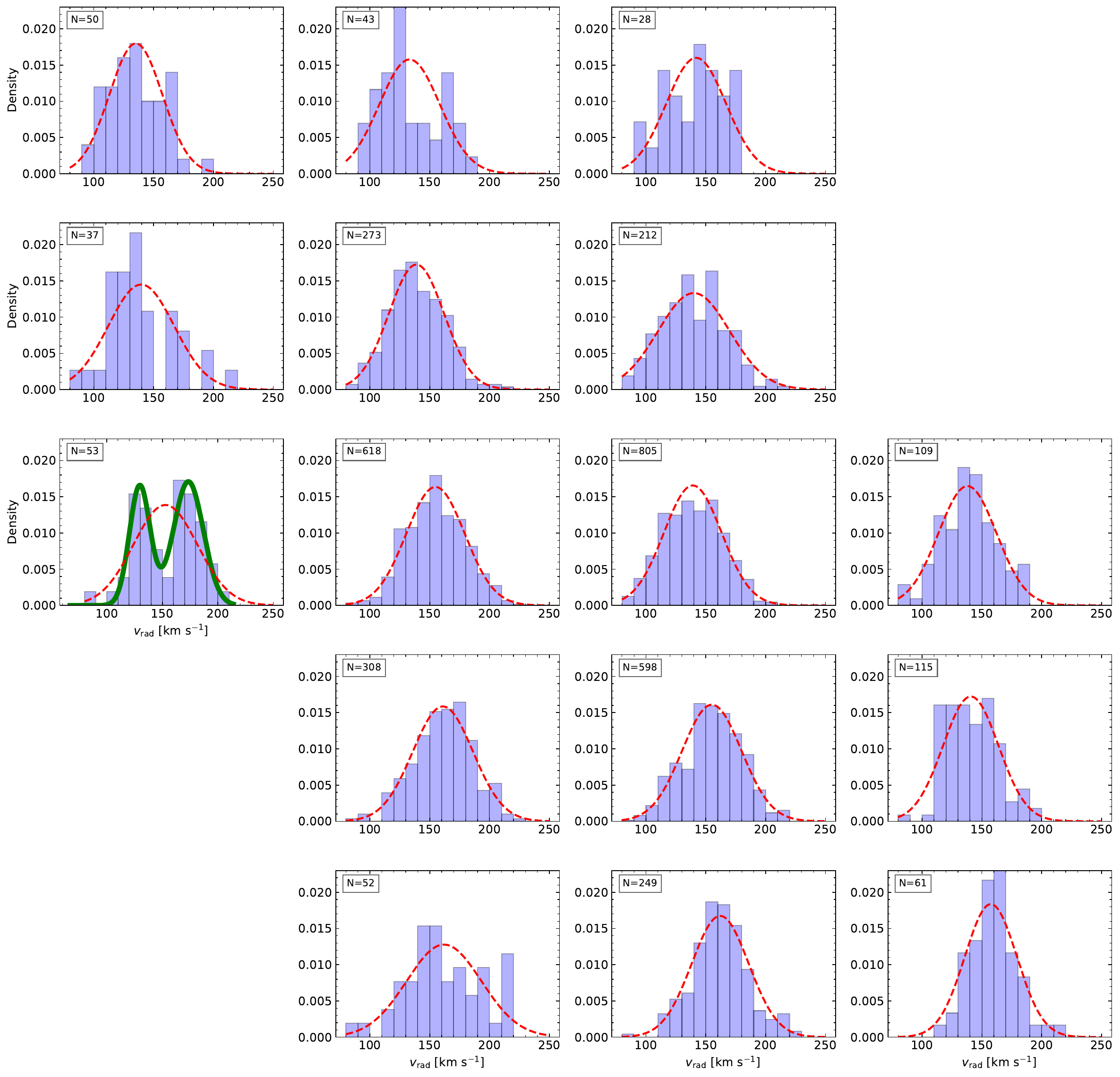}
    \caption{Like Fig.\ref{fig:fields_dis} but for the radial velocities of the HEALPix subsamples.}    
    \label{fig:fields_vr}
\end{figure*}
\begin{figure*}
    \centering
    \includegraphics[width=\textwidth]{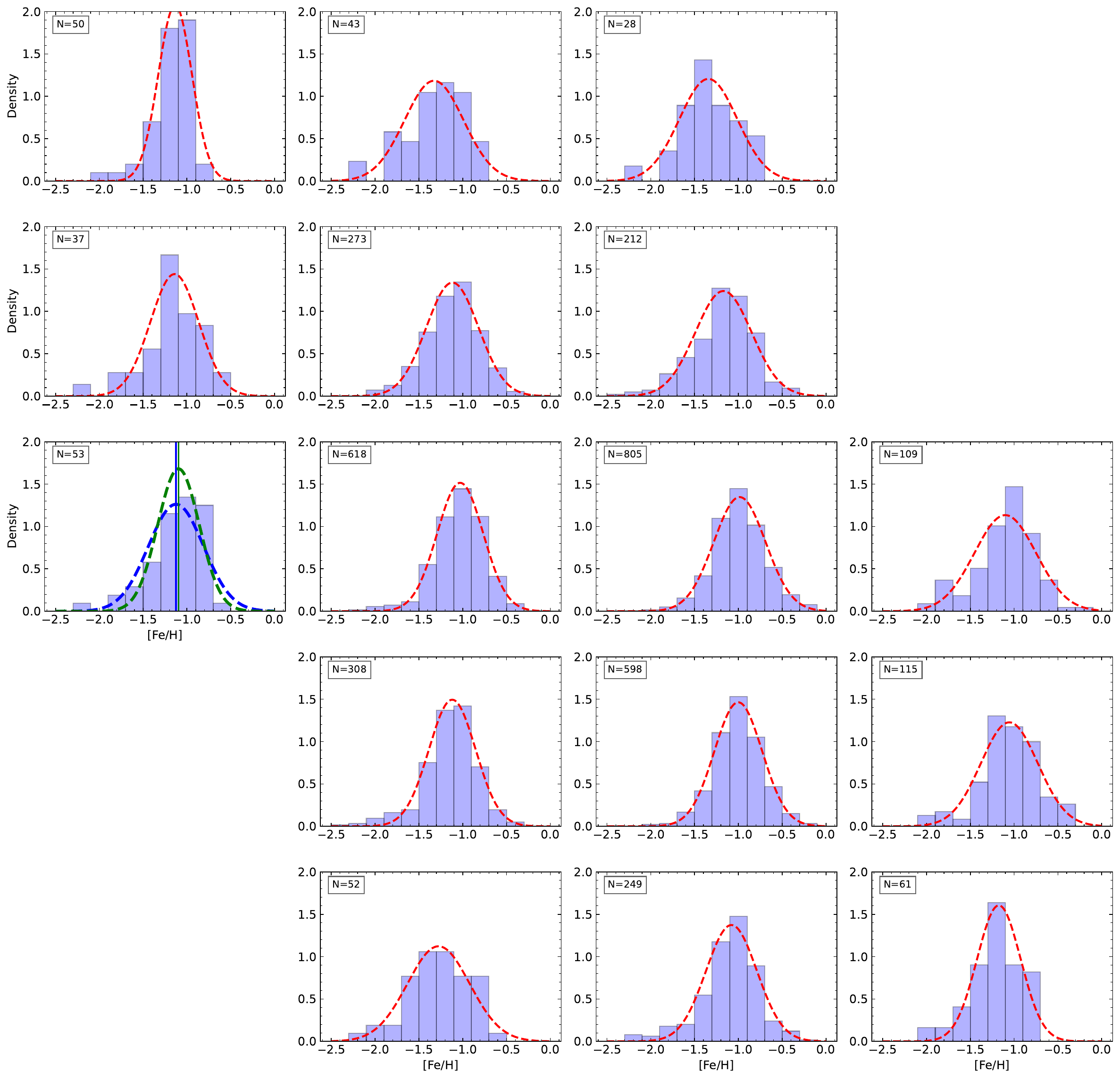}
    \caption{Like Fig.\ref{fig:fields_dis}, but for the metallicity of the HEALPix subsamples. The eastern region does not show two separate components; the green and blue dashed profiles correspond to two $v_{\rm rad}$  components in the same subsample of Fig.\ref{fig:fields_vr}.}  
    \label{fig:fields_metal}
\end{figure*}
\unskip
\input{Tables/healpix}
\unskip

\subsection{Decoupling of Kinematics and Chemistry in the Eastern SMC} 
\label{sec:3d}
Motivated by the extensive spatial coverage and high sample size of our spectroscopic survey, we explored the SMC's structural complexity by dividing the dataset into spatial subsamples using the HEALPix\footnote{http://healpix.sourceforge.net} pixelization scheme \citep{2005ApJ...622..759G,Zonca2019}. This allowed for uniform sky partitioning and analysis of variations in three dimensions: distance, velocity, and metallicity. Each HEALPix cell contains more than twenty stars, and we focus our analysis on 16 representative sub-regions across a range of Right Ascensions (RA) and Declinations (Dec).

To probe the three-dimensional nature of the SMC, we estimate distances ($d$) using \textit{Gaia} EDR3 proper motions and the relation from \citet{Almeida2024}:
\[
d = \frac{\nu_{\mathrm{tan}}}{4.74 \, \mu}
\]
where $\mu$ is the total \textit{Gaia} proper motion (in\,mas\,yr$^{-1}$) and $\nu_{\mathrm{tan}}$ is the systematic tangential velocity of the SMC, adopted here as 398\,km\,s$^{-1}$. 

With this method, we accept those distances ranging from 30 to 90\,kpc. The distributions of distance, velocity, and metallicity for each subsample are shown in Fig.~\ref{fig:fields_dis}, \ref{fig:fields_vr}, and \ref{fig:fields_metal}, respectively, with summary statistics provided in Table~\ref{tab:fields}. 
Figure~\ref{fig:fields_dis} reveals a pronounced spatial variation in line of sight depth. The eastern SMC exhibits closer mean distances (down to $\sim$50\,kpc), while the western regions extend out to nearly 60\,kpc. Notably, the eastern subsample shows a clear bimodal distance distribution, with two Gaussian components centred at approximately 47 and 56\,kpc, respectively. The observed distance bimodality is consistent with previous findings \citep{2013Nidever, Almeida2024} and has been interpreted as a consequence of a past interaction between the LMC and SMC. Based on the simulations, the foreground structure is thought to represent a tidal extension of the SMC, likely formed through the stripping of material from its disk \citep{2012ApJ...750...36D}. This scenario aligns with suggestions that both the stellar substructures and the gaseous features associated with the Magellanic Clouds originate from tidal forces induced during their mutual interaction.

Figure~\ref{fig:fields_vr} supports this interpretation by revealing corresponding bimodal structures in radial velocity. In the eastern sub-regions with double-peaked distances, we also observe two $v_{\rm rad}$  components at 129 and 173\,km\,s$^{-1}$, suggesting kinematically distinct populations. These findings strengthen the case for the existence of two separate stellar components in the eastern SMC—possibly a foreground structure and a more extended background population—both shaped by dynamical interactions.

Figure~\ref{fig:fields_metal} shows that the mean metallicity across the subsamples remains relatively uniform, ranging from [Fe/H]~$\sim -1.1$\,dex near the centre to $\sim -1.3$\,dex at the edges. Importantly, despite the spatial and kinematic substructures in the east, the metallicity distributions do not show significant bimodality. Gaussian fits to velocity-selected subsets ($V_r > 150$ vs. $V_r < 150$\,km\,s$^{-1}$) reveal nearly identical peak metallicities, suggesting no strong chemical differentiation between the kinematic components. This result is in agreement with \citet{Almeida2024}, who also found that eastern near and far sub-populations share similar enrichment histories. The same result is obtained if the global metallicity gradient is removed from the individual [Fe/H] values.

Overall, our three-dimensional analysis shows the eastern SMC's structural bifurcation in distance and velocity is not mirrored by a chemical one. Within our measurement precision, these components are chemically homogeneous, suggesting a common origin rather than the accretion of an external system. These features are therefore more likely attributable to internal dynamics driven by Magellanic interactions, as predicted by simulations. However, a definitive confirmation awaits more precise abundance analyses incorporating a wider suite of elemental tracers.

\begin{figure}
     \centering
    \includegraphics[width=\linewidth]{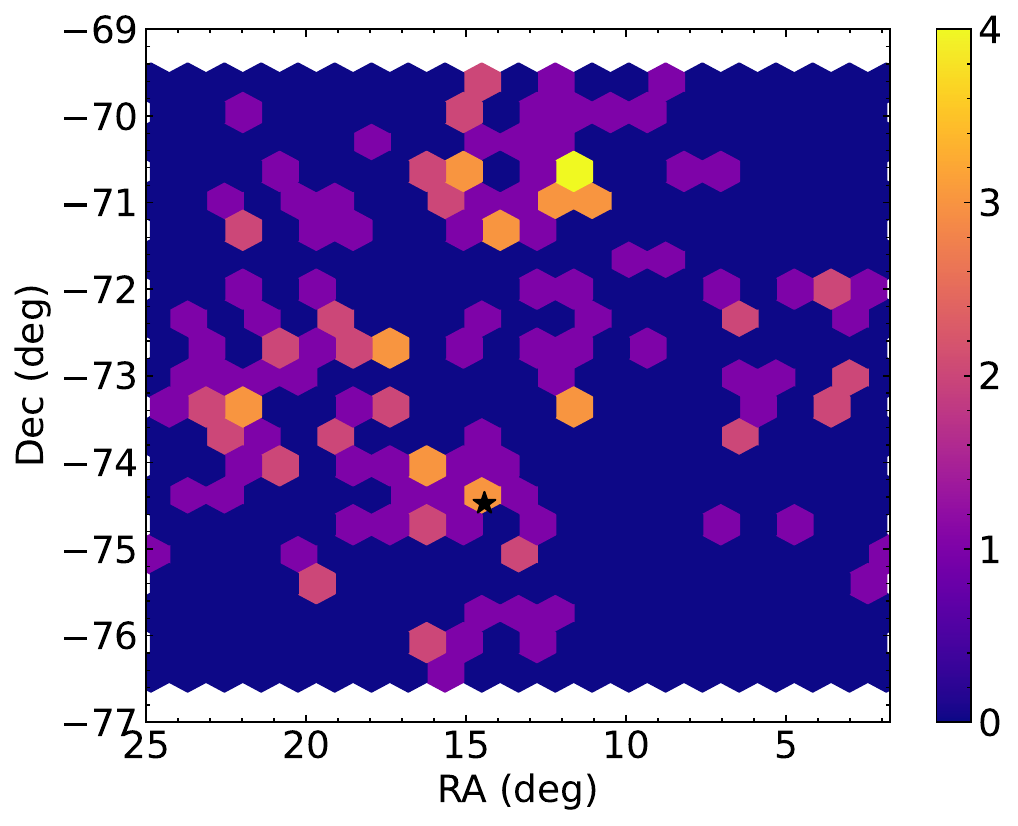}
             \caption{Map of metal-poor stars with [Fe/H]$\leq$-1.7\,dex, where the colour bar represents stellar number across different regions. The black star shows one region corresponding to the NGC~339 globular cluster.}
    \label{fig:metal_poor}
\end{figure}

\subsection{Metal-poor Map}
\label{sec:mpm}

Our sample includes 161 stars with [Fe/H]~$< -1.7$\,dex, several of which are classified as very metal-poor ([Fe/H]~$< -2$\,dex), including one extremely metal-poor object ([Fe/H]~$< -3$\,dex). Given the role of metal-poor stars as tracers of early star formation and chemical evolution, we examine their spatial and kinematic properties to assess whether they represent a distinct population compared to the dominant SMC field stars.
To investigate whether these stars may be associated with known globular clusters, we cross-matched our sample with the catalogue of \citet{2020Bica}. Only one object, NGC~339, an intermediate-age cluster ($6 \pm 0.5$\,Gyr; \citealt{Glatt2008}), aligns spatially with one of these metal-poor stars. Recent high-resolution spectroscopy by \citet{2023Mucciarelli} provides a metallicity of [Fe/H]~$= -1.24 \pm 0.02$\,dex, consistent with the SMC average and significantly more metal-rich than the stars in our metal-poor sample. Thus, we conclude that our metal-poor stars are not associated with known clusters, including NGC~339.

Figure~\ref{fig:metal_poor} displays the spatial distribution of these metal-poor stars. They are found throughout the SMC, but with a relative enhancement in the outer regions—particularly towards the east. It is important to clarify the expectations of the outside-in formation scenario: it predicts that metal-poor stars, being older, should be widespread across the galaxy, while metal-rich stars, formed later, are expected to be centrally concentrated (e.g., \citealt{Hidalgo2013}). Thus, the presence of metal-poor stars in both the inner and outer SMC is consistent with this scenario. The spatial asymmetry we observe, notably the eastern enhancement, may reflect tidal effects from the LMC rather than intrinsic chemical gradients. Indeed, \citet{2013Nidever} demonstrate that the SMC has a tidally stretched stellar component extending eastward towards the Magellanic Bridge. More recently, \citet{Almeida2024} find kinematic and distance bifurcations in this region without accompanying chemical differences, suggesting a common origin that has been dynamically disturbed.

\begin{figure}
     \centering
    \includegraphics[width=\linewidth]{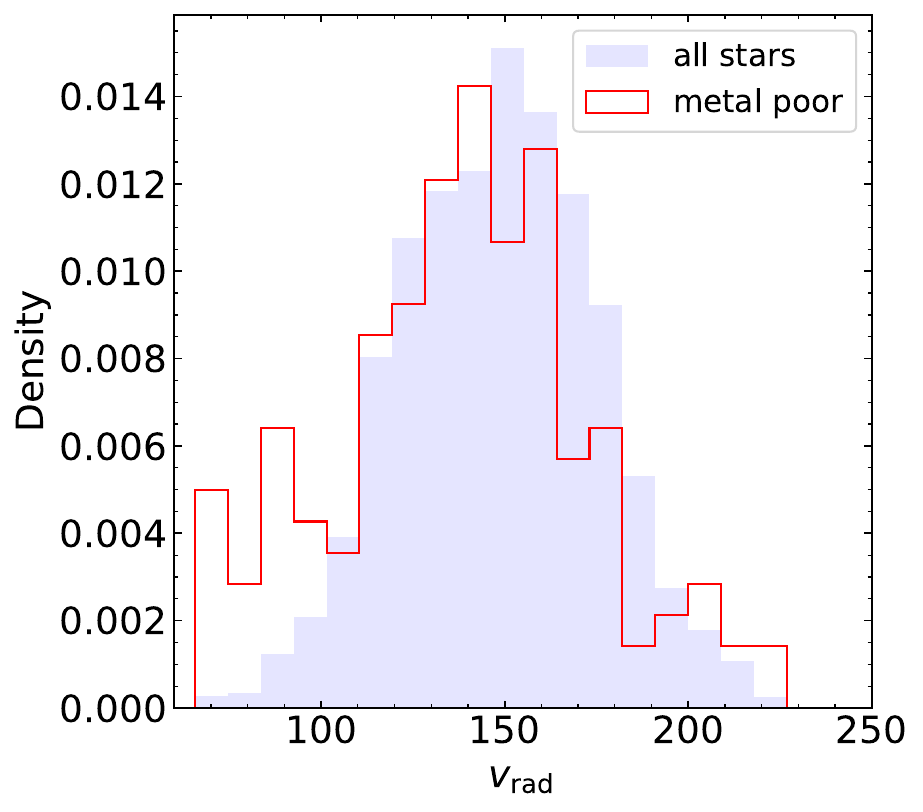}
             \caption{Histogram showing metal-poor stars with [Fe/H]$\leq$-1.7\,dex in red, compared to the full SMC stellar population in filled blue.}
    \label{fig:metal_poor_hist}
\end{figure}

To assess whether the metal-poor stars are kinematically distinct, we compare their radial velocity distribution to that of the full RGB sample in Fig.~\ref{fig:metal_poor_hist}, showing metal-poor stars in red and the full sample in blue shading. Both distributions broadly overlap, but a significant difference appears as an excess of metal-poor stars between 65 and 100\,km\,s$^{-1}$. In contrast, there is a notable deficit between 180 and 190\,km\,s$^{-1}$, which enhances the appearance of a peak around 210\,km\,s$^{-1}$. However, this latter feature coincides with the main RGB population and is not considered independently significant. The primary kinematic distinction is the low-velocity excess, which may hint at a distinct substructure within the metal-poor component.
Since the 65–100\,km\,s$^{-1}$ velocity range can overlap with that of Milky Way halo stars, we assessed potential foreground contamination using Gaia EDR3 proper motions. The distances we derived for these stars, as described in Sect.~\ref{sec:3d}, fall between those of the MW halo and the SMC. This overlap suggests that these stars are likely foreground contaminants.

\begin{figure}
     \centering
    \includegraphics[width=\linewidth]{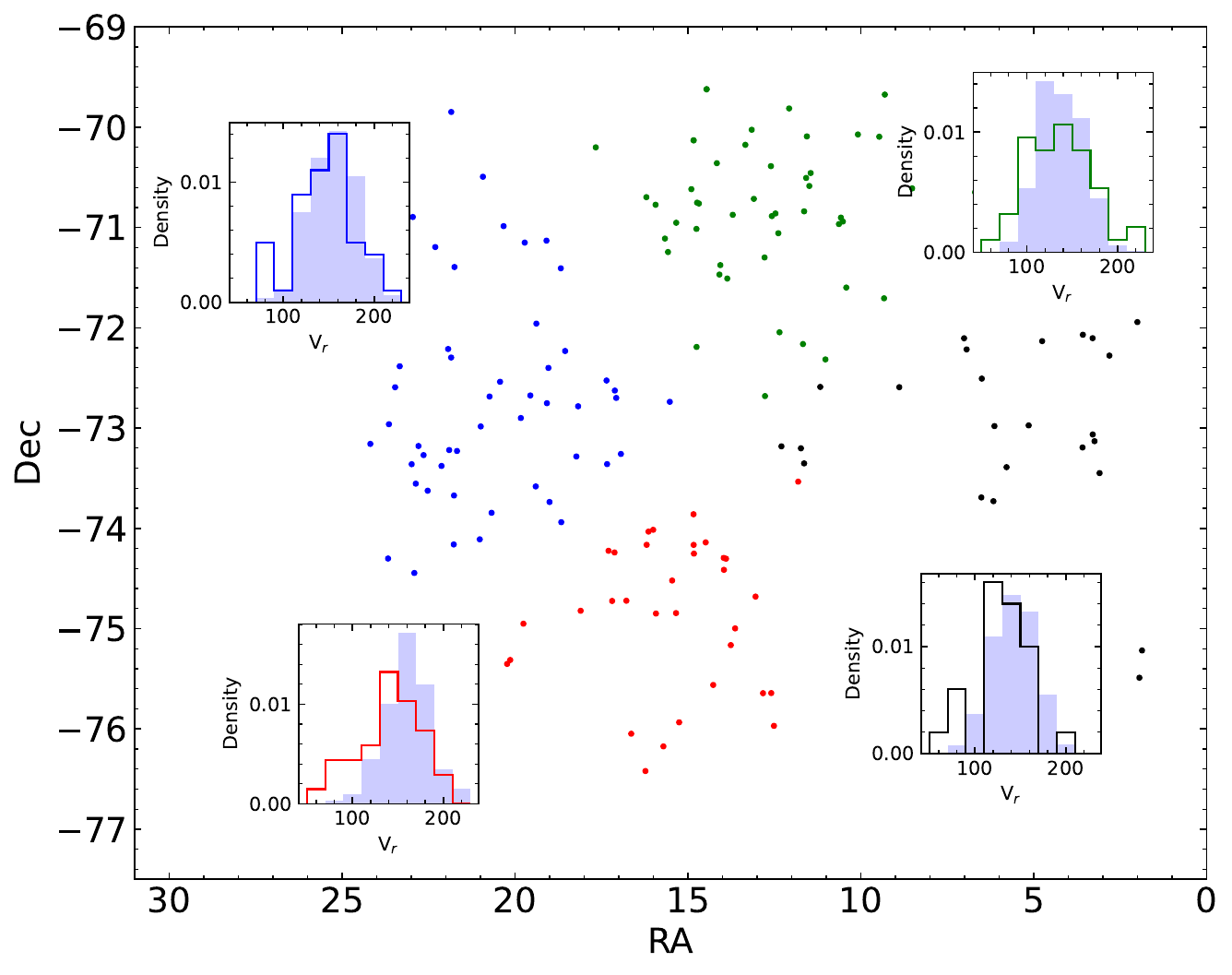}
             \caption{Map of metal-poor stars in the azimuthal quadrants, with histograms of their radial velocities ($v_{\rm rad}$ ) overlaid. The background-filled blue histograms represent the general stellar population in each region.}
    \label{fig:metal_poor_azimuth}
\end{figure}

To explore potential spatial variations within the kinematics of metal-poor stars, we obtain the radial velocity distribution in the same quadrants studied before (Fig.~\ref{fig:metal_poor_azimuth}), comparing with the full sample of each of them (blue shadow). The low radial velocity excess (65–100\,km\,s$^{-1}$) is everywhere because all regions show an excess of low-$v_{\rm rad}$  stars compared to the whole population. This reinforces our hypothesis that they are foreground MW contamination. However, the peak around 210\,km\,s$^{-1}$ can be seen just in the Northern and Western regions (green and black histograms). These high radial velocity stars with values well above the main peak may belong to a dynamically distinct substructure, such as an early accreted component or a stellar stream resulting from SMC-LMC interaction.

Overall, we find that the metal-poor stars are not confined to clusters and are more frequent in the SMC outskirts—particularly toward the east. While we initially identified an excess of metal-poor stars with radial velocities between 65 and 100\,km\,s$^{-1}$, further analysis suggests that these are likely foreground contaminants from the Milky Way. Excluding this velocity range, the spatial and kinematic distribution of the remaining metal-poor stars is broadly consistent with an outside-in formation scenario. The spatial asymmetries—especially the eastern concentration— are likely the result of tidal interactions with the LMC, which have reshaped the SMC’s outer stellar populations without introducing clear chemical discontinuities.


\unskip
\input{Tables/depth}
\unskip
\subsection{Impact of line of sight depth}

In this section, we assess whether variations in the line of sight depth of the SMC could significantly influence our derived metallicities and, by extension, our results. The SMC is known to exhibit a substantial line of sight depth, ranging from a few kiloparsecs to up to $\sim$14\,kpc depending on stellar population and location, as estimated using red clump stars by \citet{2012Subramanian}. This corresponds to a dispersion in distance modulus between 0.03 to 0.46\,mag, though for the purpose of this analysis we adopt a conservative average depth variation of 3–4\,kpc, or $\pm$0.14\,mag in distance modulus.

We recomputed the metallicities using adjusted distance moduli, shifting the absolute magnitudes by $\pm$0.14\,mag and recalculating [Fe/H] values accordingly. The results are presented in Table~\ref{tab:depth}. These adjusted distributions show no significant change in the overall metallicity: the median shifts by less than 0.05\,dex and the dispersion remains within the original uncertainty.

To visualise any potential spatial impact, we regenerated the metal-poor map using the updated metallicities. Applying a reduction of 0.14\,mag in the distance modulus results in a higher median metallicity of [Fe/H]~$= -1.03 \pm 0.01$\,dex, decreasing the number of stars falling below our metal-poor threshold. Consequently, regions originally containing a single metal-poor star may fall below detection, though denser groupings, with two or three stars, persist, albeit with reduced densities. The highest-density region, four stars, remains unchanged.

Conversely, increasing the distance modulus by 0.14\,mag lowers the median metallicity to [Fe/H]~$= -1.08 \pm 0.01$\,dex, leading to an increase in the number of stars classified as metal-poor. This creates additional low-density groupings across the map, including two regions with four stars—one in the north and one in the south. Notably, no new metal-poor concentrations appear in the western quadrant. The effect of this shift is to make the northern SMC appear slightly more metal-poor than in the original map.

Overall, the spatial distribution of metal-poor stars is only marginally affected by plausible line of sight depth variations. High-density features remain in place, and the global trends, including the eastern enhancement, are preserved. We conclude that the line of sight depth does not significantly impact our metallicity determinations or the resulting interpretation of the metal-poor population.

\section{Conclusions}
\label{sec:conclusions}

This work presents the most spatially extended and homogeneous spectroscopic study of field stars in the Small Magellanic Cloud (SMC) to date. Using infrared Ca\,\textsc{ii} triplet (CaT) spectroscopy of 3,697 red giant stars from AAOmega, we have constructed a two-dimensional metallicity map and investigated the chemical and kinematic structure of the galaxy.

Our results provide new insights into the formation and dynamical evolution of the SMC, which we summarise below:

\begin{itemize}
    \item \textbf{Outside-in formation and central inversion:} We confirm a negative metallicity gradient ($-0.064 \pm 0.007$\,dex\,deg$^{-1}$) out to 6\fdg0, consistent with an outside-in chemical evolution scenario. For the first time, we detect hints of a positive metallicity gradient within 1\fdg2, likely reflecting radial migration or centralised chemical enrichment.

    \item \textbf{Azimuthal metallicity asymmetries:} We find flatter gradients in the eastern and southern quadrants and steeper ones in the north and west. These asymmetries are consistent with tidal interaction effects from the Large Magellanic Cloud (LMC), which appear to have redistributed stars and altered the chemical structure of the SMC.

    \item \textbf{Coherent structure in the eastern SMC:} Despite clear distance and velocity bifurcations in the east, no corresponding metallicity differences are found. This supports the view that the near and far eastern structures share a common chemical origin, in agreement with recent findings from \citet{Almeida2024}.

    \item \textbf{Metal-poor population structure:} We identify 161 stars with [Fe/H]~$< -1.7$\,dex, mostly located in the outskirts—especially the eastern SMC. 
    One extremely metal-poor star ([Fe/H]~$< -3$\,dex) is also found. These stars are not associated with known clusters and include a kinematically distinct high-velocity component that may represent an early substructure.

    \item \textbf{Robustness to depth effects:} Shifting the distance modulus by $\pm$0.14\,mag does not significantly change the median [Fe/H] or affect the location of high-density metal-poor regions, confirming that our results are robust against line of sight depth variations.
\end{itemize}

In summary, the SMC exhibits a rich and complex chemical structure shaped by both internal evolution and external tidal influences. The combination of metallicity gradients, asymmetries, and outer-halo substructure paints a picture of a dwarf galaxy in active transformation, retaining signatures of its early star formation history while responding to its interaction with the LMC.

\section*{Acknowledgements}

This paper includes data that has been provided by AAO Data Central  (datacentral.org.au). Based on data acquired through the Australian Astronomical Observatory, under programs A/2011B/08 and A/2011B/13. The research leading to these results has received funding from the European Community's Seventh Framework Programme (FP7/2007-2013) under grant agreement number RG226604 (OPTICON).
MDL acknowledges financial support from the project \textquote{LEGO – Reconstructing the building blocks of the Galaxy by chemical tagging} (PI: Mucciarelli) granted by the Italian MUR through contract PRIN2022LLP8TK\_001.

\section*{Data availability}

The data underlying this article are available from the Anglo-Australian Telescope data archive (https://datacentral.org.au/) and the Gaia EDR3 catalogue (https://gea.esac.esa.int/archive/). Processed spectra generated during this study will be shared upon reasonable request to the corresponding author. The measurements of the strengths of the lines will be published through CDS.

\bibliographystyle{mnras}
\bibliography{biblio}

\appendix

\label{lastpage}
\end{document}

%% file: Tables/Table_validation.tex
\begin{table*}
    \caption{The distance modulus, reddening, and reference [Fe/H] values for the four clusters are listed. The CaT metallicities for these clusters are derived from absolute magnitudes in four bands: $K$, $G$, $V$, and $I$. The column labeled N$_{K,G,V,I}$ indicates the number of stars observed in each respective band.}
    \centering
    \begin{tabular}{lccccccccccc}
          Cluster & (m - M)$_0$ &  E(B-V) & Ref. & [Fe/H]$_{ref}$ & Ref. & [Fe/H] & [Fe/H] & [Fe/H] & [Fe/H] & N$_{K,G,V,I}$  \\
          Name &mag & mag&  & dex& & $K$ band & $G$ band & $V$ band & $I$ band & &\\
		\hline
         NGC 7099 & 14.50$\pm$0.21 & 0.05$\pm$0.02  &1& -2.33$\pm$0.02  & 5 & -2.34$\pm$0.14 &-2.33$\pm$0.13& 
         -2.31$\pm$0.13& -2.36 $\pm$  0.14  & 22,22,22,22\\
         NGC 362 & 14.77$\pm$0.01  &0.03$\pm$0.01 & 2 &  -1.09$\pm$0.01& 6 & -1.02$\pm$0.16 &-1.08$\pm$0.17& -1.07 $\pm$0.09 & -1.07$\pm$0.16 &13,12,11,12\\
         NGC 104 & 13.21$\pm$0.09   & 0.03$\pm$0.01 &3  &  -0.76$\pm$0.02& C13 & -0.78$\pm$0.20 &-0.79$\pm$0.19& -0.75$\pm$0.23 & -0.75$\pm$0.24&36,36,31,31\\
         Melotte 66& 13.40$\pm$0.30 & 0.15$\pm$0.02  &4   & -0.33$\pm$0.03& 4 & -0.49$\pm$0.14 &-0.52$\pm$0.12& -0.56$\pm$0.12& -0.52$\pm$0.13 &34,34,26,25\\ 
 		\hline
    \end{tabular}
    \\References: (1) \cite{2010Smitka}; (2) \cite{2021Gontcharov}; (3) \cite{2017Brogaard};  (4) \cite{2009CarrettaC};  (5) \cite{2024Gontcharov};   (6)  \cite{2022Vargas}
    \label{tab:fe}
\end{table*}

%% file: Tables/metal_distribution.tex
\begin{table}
    \centering
    \caption{Metallicity distributions of SMC field stars derived in this study. The values include the median metallicity [Fe/H],  10th and 90th percentiles (P10, P90), and the standard deviation ($\sigma$).}
    \label{tab:metallicity_distributions}
    \begin{tabular}{lccc}
        \hline
        \textbf{Sample} & \textbf{<[Fe/H]> (P10, P90)} & \textbf{$\sigma$} \\
        \hline
        $K_s$-band &-1.05$\pm$0.01 (-1.46, -0.70) & 0.33 \\
        $G$-band & -1.03$\pm$0.01 (-1.41, -0.70) & 0.31 \\
        \hline
        \textbf{Azimuthal Regions} & & \\
East& -1.07$\pm$0.01 (-1.46, -0.74) & 0.33\\
 South& -1.09$\pm$0.02 (-1.59, -0.79) & 0.35\\
 West& -1.04$\pm$0.02 (-1.48, -0.59) & 0.35\\
 North& -1.12$\pm$0.02 (-1.62, -0.76) & 0.34\\
         \hline
    \end{tabular}
\end{table}


%% file: Tables/metal_literature.tex
\begin{table}
    \centering
    \caption{Comparison of median metallicities ([Fe/H]) and standard deviations ($\sigma$) from previous studies. Where available, uncertainties are included.}
    \label{tab:literature_comparison}
    \begin{tabular}{lcc}
        \hline
        \textbf{Study} & \textbf{<[Fe/H]>} & \textbf{$\sigma$} \\
        \hline
         \citet{Carrera2008} & $\sim -1$ & --- \\
        \citet{Parisi2010} & $-1.00 \pm 0.02$ & $0.30 \pm 0.01$ \\
        \citet{2014dobbie}  & $-0.99 \pm 0.02$ & --- \\
        \citet{Parisi2016}  & $-0.97 \pm 0.01$ & $0.30 \pm 0.01$ \\
        \citet{Bortoli2022}& $\sim -1$ & 0.3\\
        \citet{mucciarelli2023} & $\sim -1$ & --- \\
        \hline
    \end{tabular}
\end{table}

%% file: Tables/radial.tex
\begin{table}
	\centering
    \renewcommand{\arraystretch}{1.2} 
	\caption{Statistics of the metallicity distribution in each elliptical annulus studied as a function of distance with the standard deviation in each bin.}
	\label{tab:radial}
	\begin{tabular}{lcccc}
    \hline
	$d$ & N$_{tot}$ & <[Fe/H]> (P10, P90)& $\sigma$ \\
		\hline
0\fdg3 & 49 & -1.02 $\pm$ 0.06 (-1.44, -0.76) & 0.3 \\
0\fdg6 & 178 & -0.99 $\pm$ 0.03 (-1.32, -0.66) & 0.3 \\
0\fdg9 & 294 & -0.99 $\pm$ 0.01 (-1.34, -0.65) & 0.3 \\
1\fdg2 & 338 & -0.97 $\pm$ 0.02 (-1.30, -0.60) & 0.3 \\
1\fdg5 & 405 & -1.03 $\pm$ 0.02 (-1.41, -0.63) & 0.3 \\
1\fdg8 & 392 & -1.00 $\pm$ 0.02 (-1.36, -0.64) & 0.3 \\
2\fdg1 & 377 & -1.05 $\pm$ 0.02 (-1.39, -0.68) & 0.3 \\
2\fdg4 & 292 & -1.05 $\pm$ 0.02 (-1.44, -0.73) & 0.3 \\
2\fdg7 & 251 & -1.08 $\pm$ 0.02 (-1.59, -0.76) & 0.3 \\
3\fdg0 & 218 & -1.08 $\pm$ 0.03 (-1.66, -0.73) & 0.4 \\
3\fdg3 & 208 & -1.18 $\pm$ 0.02 (-1.59, -0.82) & 0.3 \\
3\fdg6 & 176 & -1.12 $\pm$ 0.02 (-1.63, -0.76) & 0.4 \\
3\fdg9 & 178 & -1.15 $\pm$ 0.02 (-1.72, -0.81) & 0.4 \\
4\fdg2 & 132 & -1.10 $\pm$ 0.03 (-1.58, -0.82) & 0.3 \\
4\fdg5 & 73 & -1.21 $\pm$ 0.04 (-1.69, -0.88) & 0.4 \\
4\fdg8 & 46 & -1.24 $\pm$ 0.08 (-2.04, -0.92) & 0.4 \\
5\fdg1 & 33 & -1.24 $\pm$ 0.08 (-1.61, -0.78) & 0.3 \\
5\fdg4 & 10 & -1.37 $\pm$ 0.12 (-1.72, -1.14) & 0.3 \\
5\fdg7 & 4 & -1.52 $\pm$ 0.19 (-1.86, -1.21) & 0.3 \\
6\fdg0 & 3 & -1.23 $\pm$ 0.24 (-1.40, -0.88) & 0.3 \\
\hline
	\end{tabular}
\end{table}

%% file: Tables/MG.tex
\begin{table}
	\centering
    \renewcommand{\arraystretch}{1.2} 
    \setlength{\tabcolsep}{6mm}
	\caption{The metallicity gradient of the azimuthal regions of the SMC.}
	\label{tab:mg}
	\begin{tabular}{lccc}
    \hline
		 Region & Range [deg] & Metallicity gradients [dex\,deg$^{-1}$] \\
    \hline
All & $<$1.2 & 0.06$\pm$0.05 \\
All & 1.2-5.0& -0.06$\pm$0.01 \\
\hline
West  & \multirow{4}{*}{1.2-3.8} &  -0.09$\pm$0.02 \\
North &  &  -0.08$\pm$0.02 \\
East  &   & -0.05$\pm$0.01 \\
South &  & -0.04$\pm$0.01\\
            \hline
	\end{tabular}
\end{table}

%% file: Tables/azimuth.tex
\begin{table}
	\centering
    \renewcommand{\arraystretch}{1.5} 
	\caption{The number of stars contributing to each radial bin per quadrant. The value of $d$ represents the average of the radial measurements across the different quadrants.}
	\label{tab:azimuth}
	\begin{tabular}{lccccc}
    \hline 
	$d$ & East &  South&  West&  North \\
		\hline
1\fdg3 & 2 & 0 & 3 & 1\\
1\fdg8 & 188 & 70 & 97 &140\\
2\fdg1 & 120 & 94 & 86 &116\\
2\fdg4 & 92 & 78 & 66 &97\\
2\fdg8 & 85 & 64 & 44 &83\\
3\fdg1 & 105 & 39 & 46& 62\\
3\fdg5 & 97 & 42 & 39 & 50\\
3\fdg8 & 57 & 44 & 33 &41\\
4\fdg2 & 34 & 28 & 0 &34\\
4\fdg5 & 11 & 15 & 0 &17\\
4\fdg8 & 11 & 9 & 0 &18\\
5\fdg1 & 5 & 1 & 0 &3\\
5\fdg5 & 0 & 1 & 0 &4\\
5\fdg9 & 0 & 0 & 0 &3\\
\hline
	\end{tabular}
\end{table}

%% file: Tables/healpix.tex
\begin{table}
	\centering
    \renewcommand{\arraystretch}{1.2} 
	\caption{A table summarizing the properties of histograms for the subsamples in Fig. \ref{fig:fields_dis}, \ref{fig:fields_vr} and \ref{fig:fields_metal}.}       
	\label{tab:fields}
	\begin{tabular}{ccccc}
		 &  & & & \\
		\hline
        \multirow{5}{*}{\shortstack{(RA,Dec)\\(deg,deg)}}&21.9,-69.4 & 14.9,-69.4 & 9.4,-69.4 &\\
        &22.6,-70.9 & 16.3,-70.9  & 10.8,-70.9 &\\
        &23.8,-72.4 & 17.8,-72.4  & 11.4,-72.4 &4.1,-72.4\\
        & & 20.3,-73.9  & 12.9,-73.9 &5.9,-73.9\\
        & & 21.3,-75.3  & 14.4,-75.3  &5.8,-75.3 \\
                 \hline
         \multirow{5}{*}{\shortstack{$<d>$\\(kpc)}}& 55$\pm$6& 58$\pm$8&61$\pm$6\\
         & 53$\pm$7 &58$\pm$5  &60$\pm$6\\
         & 52$\pm$5& 58$\pm$6  &60$\pm$5&61$\pm$6\\
         & &56$\pm$7& 60$\pm$6 &62$\pm$7\\
         & &53$\pm$7& 59$\pm$6 &62$\pm$5\\
                  \hline
         \multirow{5}{*}{\shortstack{<$v_{\rm rad}$> \\(\,km\,s$^{-1}$)}}
  & 135$\pm$22& 133$\pm$25 &142$\pm$25\\
         & 140$\pm$27& 139$\pm$23 &140$\pm$29\\
         & 152$\pm$28& 155$\pm$24 &139$\pm$24&138$\pm$24\\
         & &161$\pm$25& 155$\pm$25 &141$\pm$23\\
         & &162$\pm$31& 162$\pm$24 &157$\pm$21\\
                  \hline
         \multirow{5}{*}{\shortstack{<[Fe/H]>\\(dex)}}& -1.14$\pm$0.19& -1.33$\pm$0.34   &-1.34$\pm$0.33\\
         & -1.14$\pm$0.28& -1.12$\pm$0.30  &-1.17$\pm$0.32\\
         & -1.11$\pm$0.28& -1.03$\pm$0.26  &-0.99$\pm$0.30&-1.10$\pm$0.35\\
         & &-1.12$\pm$0.27& -1.00$\pm$0.27 &-1.05$\pm$0.33 \\ 
         & &-1.28$\pm$0.36& -1.08$\pm$0.29 &-1.17$\pm$0.25\\
         \hline
	\end{tabular}
\end{table}

%% file: Tables/depth.tex
\begin{table}
	\centering
	\caption{Properties of metallicity distributions after applying $\pm$0.14 mag on the distance modulus ($\mu$).}
	\label{tab:depth}
	\begin{tabular}{lccc}
		 Distribution & [Fe/H](Q1,Q3) & $\sigma$ \\
		\hline
         $K_s$-band ($\mu-0.14$) & -1.03$\pm$0.01 (-1.22, -0.84) & 0.34\\
$G$-band ($\mu-0.14$) & -1.00$\pm$0.01 (-1.18, -0.83) & 0.31\\
        		\hline
$K_s$-band ($\mu+0.14$) & -1.08$\pm$0.01 (-1.27, -0.89) & 0.33\\
$G$-band ($\mu+0.14$) & -1.05$\pm$0.01 (-1.23, -0.88) & 0.30\\
\hline
	\end{tabular}
\end{table}

%% file: paper_draft.bbl
\begin{thebibliography}{}
\makeatletter
\relax
\def\mn@urlcharsother{\let\do\@makeother \do\$\do\&\do\#\do\^\do\_\do\%\do\~}
\def\mn@doi{\begingroup\mn@urlcharsother \@ifnextchar [ {\mn@doi@} {\mn@doi@[]}}
\def\mn@doi@[#1]#2{\def\@tempa{#1}\ifx\@tempa\@empty \href {http://dx.doi.org/#2} {doi:#2}\else \href {http://dx.doi.org/#2} {#1}\fi \endgroup}
\def\mn@eprint#1#2{\mn@eprint@#1:#2::\@nil}
\def\mn@eprint@arXiv#1{\href {http://arxiv.org/abs/#1} {{\tt arXiv:#1}}}
\def\mn@eprint@dblp#1{\href {http://dblp.uni-trier.de/rec/bibtex/#1.xml} {dblp:#1}}
\def\mn@eprint@#1:#2:#3:#4\@nil{\def\@tempa {#1}\def\@tempb {#2}\def\@tempc {#3}\ifx \@tempc \@empty \let \@tempc \@tempb \let \@tempb \@tempa \fi \ifx \@tempb \@empty \def\@tempb {arXiv}\fi \@ifundefined {mn@eprint@\@tempb}{\@tempb:\@tempc}{\expandafter \expandafter \csname mn@eprint@\@tempb\endcsname \expandafter{\@tempc}}}

\bibitem[\protect\citeauthoryear{{Almeida} et~al.,}{{Almeida} et~al.}{2024}]{Almeida2024}
{Almeida} A.,  et~al., 2024, \mn@doi [\mnras] {10.1093/mnras/stae373}, \href {https://ui.adsabs.harvard.edu/abs/2024MNRAS.529.3858A} {529, 3858}

\bibitem[\protect\citeauthoryear{{Aparicio} \& {Tikhonov}}{{Aparicio} \& {Tikhonov}}{2000}]{apariciotikhonov2000}
{Aparicio} A.,  {Tikhonov} N.,  2000, \mn@doi [\aj] {10.1086/301360}, \href {https://ui.adsabs.harvard.edu/abs/2000AJ....119.2183A} {119, 2183}

\bibitem[\protect\citeauthoryear{{Astropy Collaboration} \& et al.}{{Astropy Collaboration} \& et~al.}{2013}]{2013Astropy}
{Astropy Collaboration} et al. 2013, \mn@doi [\aap] {10.1051/0004-6361/201322068}, \href {https://ui.adsabs.harvard.edu/abs/2013A&A...558A..33A} {558, A33}

\bibitem[\protect\citeauthoryear{{Astropy Collaboration} \& et al.}{{Astropy Collaboration} \& et~al.}{2018}]{2018Astropy}
{Astropy Collaboration} et al. 2018, \mn@doi [\aj] {10.3847/1538-3881/aabc4f}, \href {https://ui.adsabs.harvard.edu/abs/2018AJ....156..123A} {156, 123}

\bibitem[\protect\citeauthoryear{{Barker}, {Sarajedini}, {Geisler}, {Harding}  \& {Schommer}}{{Barker} et~al.}{2007}]{2007Barker}
{Barker} M.~K.,  {Sarajedini} A.,  {Geisler} D.,  {Harding} P.,   {Schommer} R.,  2007, \mn@doi [\aj] {10.1086/511186}, \href {https://ui.adsabs.harvard.edu/abs/2007AJ....133.1138B} {133, 1138}

\bibitem[\protect\citeauthoryear{{Barker}, {Ferguson}, {Cole}, {Ibata}, {Irwin}, {Lewis}, {Smecker-Hane}  \& {Tanvir}}{{Barker} et~al.}{2011}]{2011Barker}
{Barker} M.~K.,  {Ferguson} A.~M.~N.,  {Cole} A.~A.,  {Ibata} R.,  {Irwin} M.,  {Lewis} G.~F.,  {Smecker-Hane} T.~A.,   {Tanvir} N.~R.,  2011, \mn@doi [\mnras] {10.1111/j.1365-2966.2010.17458.x}, \href {https://ui.adsabs.harvard.edu/abs/2011MNRAS.410..504B} {410, 504}

\bibitem[\protect\citeauthoryear{{Belokurov} \& {Erkal}}{{Belokurov} \& {Erkal}}{2019}]{2019MNRAS.482L...9B}
{Belokurov} V.~A.,  {Erkal} D.,  2019, \mn@doi [\mnras] {10.1093/mnrasl/sly178}, \href {https://ui.adsabs.harvard.edu/abs/2019MNRAS.482L...9B} {482, L9}

\bibitem[\protect\citeauthoryear{{Besla}, {Kallivayalil}, {Hernquist}, {van der Marel}, {Cox}  \& {Kere{\v{s}}}}{{Besla} et~al.}{2010}]{Besla2010}
{Besla} G.,  {Kallivayalil} N.,  {Hernquist} L.,  {van der Marel} R.~P.,  {Cox} T.~J.,   {Kere{\v{s}}} D.,  2010, \mn@doi [\apjl] {10.1088/2041-8205/721/2/L97}, \href {https://ui.adsabs.harvard.edu/abs/2010ApJ...721L..97B} {721, L97}

\bibitem[\protect\citeauthoryear{{Besla}, {Kallivayalil}, {Hernquist}, {van der Marel}, {Cox}  \& {Kere{\v{s}}}}{{Besla} et~al.}{2012}]{2012MNRAS.421.2109B}
{Besla} G.,  {Kallivayalil} N.,  {Hernquist} L.,  {van der Marel} R.~P.,  {Cox} T.~J.,   {Kere{\v{s}}} D.,  2012, \mn@doi [\mnras] {10.1111/j.1365-2966.2012.20466.x}, \href {https://ui.adsabs.harvard.edu/abs/2012MNRAS.421.2109B} {421, 2109}

\bibitem[\protect\citeauthoryear{{Bica}, {Westera}, {Kerber}, {Dias}, {Maia}, {Santos}, {Barbuy}  \& {Oliveira}}{{Bica} et~al.}{2020}]{2020Bica}
{Bica} E.,  {Westera} P.,  {Kerber} L. d.~O.,  {Dias} B.,  {Maia} F.,  {Santos} Jr. J. F.~C.,  {Barbuy} B.,   {Oliveira} R. A.~P.,  2020, \mn@doi [\aj] {10.3847/1538-3881/ab6595}, \href {https://ui.adsabs.harvard.edu/abs/2020AJ....159...82B} {159, 82}

\bibitem[\protect\citeauthoryear{{Brogaard}, {VandenBerg}, {Bedin}, {Milone}, {Thygesen}  \& {Grundahl}}{{Brogaard} et~al.}{2017}]{2017Brogaard}
{Brogaard} K.,  {VandenBerg} D.~A.,  {Bedin} L.~R.,  {Milone} A.~P.,  {Thygesen} A.,   {Grundahl} F.,  2017, \mn@doi [\mnras] {10.1093/mnras/stx378}, \href {https://ui.adsabs.harvard.edu/abs/2017MNRAS.468..645B} {468, 645}

\bibitem[\protect\citeauthoryear{{Carbajo-Hijarrubia} et~al.,}{{Carbajo-Hijarrubia} et~al.}{2024}]{2024Carbajo}
{Carbajo-Hijarrubia} J.,  et~al., 2024, \mn@doi [\aap] {10.1051/0004-6361/202347648}, \href {https://ui.adsabs.harvard.edu/abs/2024A&A...687A.239C} {687, A239}

\bibitem[\protect\citeauthoryear{{Cardelli}, {Clayton}  \& {Mathis}}{{Cardelli} et~al.}{1989}]{1989ApJ...345..245C}
{Cardelli} J.~A.,  {Clayton} G.~C.,   {Mathis} J.~S.,  1989, \mn@doi [\apj] {10.1086/167900}, \href {https://ui.adsabs.harvard.edu/abs/1989ApJ...345..245C} {345, 245}

\bibitem[\protect\citeauthoryear{{Carraro}, {de Silva}, {Monaco}, {Milone}  \& {Mateluna}}{{Carraro} et~al.}{2014}]{2014Carraro}
{Carraro} G.,  {de Silva} G.,  {Monaco} L.,  {Milone} A.~P.,   {Mateluna} R.,  2014, \mn@doi [\aap] {10.1051/0004-6361/201423714}, \href {https://ui.adsabs.harvard.edu/abs/2014A&A...566A..39C} {566, A39}

\bibitem[\protect\citeauthoryear{{Carrera}, {Gallart}, {Pancino}  \& {Zinn}}{{Carrera} et~al.}{2007}]{Carrera2007}
{Carrera} R.,  {Gallart} C.,  {Pancino} E.,   {Zinn} R.,  2007, \mn@doi [\aj] {10.1086/520803}, \href {https://ui.adsabs.harvard.edu/abs/2007AJ....134.1298C} {134, 1298}

\bibitem[\protect\citeauthoryear{{Carrera}, {Gallart}, {Aparicio}, {Costa}, {M{\'e}ndez}  \& {No{\"e}l}}{{Carrera} et~al.}{2008}]{Carrera2008}
{Carrera} R.,  {Gallart} C.,  {Aparicio} A.,  {Costa} E.,  {M{\'e}ndez} R.~A.,   {No{\"e}l} N. E.~D.,  2008, \mn@doi [\aj] {10.1088/0004-6256/136/3/1039}, \href {https://ui.adsabs.harvard.edu/abs/2008AJ....136.1039C} {136, 1039}

\bibitem[\protect\citeauthoryear{{Carrera}, {Pancino}, {Gallart}  \& {del Pino}}{{Carrera} et~al.}{2013}]{Carrera2013}
{Carrera} R.,  {Pancino} E.,  {Gallart} C.,   {del Pino} A.,  2013, \mn@doi [\mnras] {10.1093/mnras/stt1126}, \href {https://ui.adsabs.harvard.edu/abs/2013MNRAS.434.1681C} {434, 1681}

\bibitem[\protect\citeauthoryear{{Carrera}, {Rodr{\'\i}guez Espinosa}, {Casamiquela}, {Balaguer Nu{\~n}ez}, {Jordi}, {Allende Prieto}  \& {Stetson}}{{Carrera} et~al.}{2017a}]{2017Carrera}
{Carrera} R.,  {Rodr{\'\i}guez Espinosa} L.,  {Casamiquela} L.,  {Balaguer Nu{\~n}ez} L.,  {Jordi} C.,  {Allende Prieto} C.,   {Stetson} P.~B.,  2017a, \mn@doi [\mnras] {10.1093/mnras/stx1526}, \href {https://ui.adsabs.harvard.edu/abs/2017MNRAS.470.4285C} {470, 4285}

\bibitem[\protect\citeauthoryear{{Carrera}, {Rodr{\'\i}guez Espinosa}, {Casamiquela}, {Balaguer Nu{\~n}ez}, {Jordi}, {Allende Prieto}  \& {Stetson}}{{Carrera} et~al.}{2017b}]{carrera2017}
{Carrera} R.,  {Rodr{\'\i}guez Espinosa} L.,  {Casamiquela} L.,  {Balaguer Nu{\~n}ez} L.,  {Jordi} C.,  {Allende Prieto} C.,   {Stetson} P.~B.,  2017b, \mn@doi [\mnras] {10.1093/mnras/stx1526}, \href {https://ui.adsabs.harvard.edu/abs/2017MNRAS.470.4285C} {470, 4285}

\bibitem[\protect\citeauthoryear{{Carretta}, {Bragaglia}, {Gratton}, {D'Orazi}  \& {Lucatello}}{{Carretta} et~al.}{2009}]{2009CarrettaC}
{Carretta} E.,  {Bragaglia} A.,  {Gratton} R.,  {D'Orazi} V.,   {Lucatello} S.,  2009, \mn@doi [\aap] {10.1051/0004-6361/200913003}, \href {https://ui.adsabs.harvard.edu/abs/2009A&A...508..695C} {508, 695}

\bibitem[\protect\citeauthoryear{{Choudhury}, {Subramaniam}, {Cole}  \& {Sohn}}{{Choudhury} et~al.}{2018}]{2018Choudhury}
{Choudhury} S.,  {Subramaniam} A.,  {Cole} A.~A.,   {Sohn} Y.~J.,  2018, \mn@doi [\mnras] {10.1093/mnras/sty087}, \href {https://ui.adsabs.harvard.edu/abs/2018MNRAS.475.4279C} {475, 4279}

\bibitem[\protect\citeauthoryear{{Choudhury} et~al.,}{{Choudhury} et~al.}{2020}]{2020Choudhury}
{Choudhury} S.,  et~al., 2020, \mn@doi [\mnras] {10.1093/mnras/staa2140}, \href {https://ui.adsabs.harvard.edu/abs/2020MNRAS.497.3746C} {497, 3746}

\bibitem[\protect\citeauthoryear{{Cioni}}{{Cioni}}{2009}]{2009Cioni}
{Cioni} M. R.~L.,  2009, \mn@doi [\aap] {10.1051/0004-6361/200912138}, \href {https://ui.adsabs.harvard.edu/abs/2009A&A...506.1137C} {506, 1137}

\bibitem[\protect\citeauthoryear{{Cole}, {Smecker-Hane}, {Tolstoy}, {Bosler}  \& {Gallagher}}{{Cole} et~al.}{2004}]{Cole2004}
{Cole} A.~A.,  {Smecker-Hane} T.~A.,  {Tolstoy} E.,  {Bosler} T.~L.,   {Gallagher} J.~S.,  2004, \mn@doi [\mnras] {10.1111/j.1365-2966.2004.07223.x}, \href {https://ui.adsabs.harvard.edu/abs/2004MNRAS.347..367C} {347, 367}

\bibitem[\protect\citeauthoryear{{D'Onghia} \& {Fox}}{{D'Onghia} \& {Fox}}{2016}]{2016ARA&A..54..363D}
{D'Onghia} E.,  {Fox} A.~J.,  2016, \mn@doi [\araa] {10.1146/annurev-astro-081915-023251}, \href {https://ui.adsabs.harvard.edu/abs/2016ARA&A..54..363D} {54, 363}

\bibitem[\protect\citeauthoryear{{De Bortoli}, {Parisi}, {Bassino}, {Geisler}, {Dias}, {Gimeno}, {Angelo}  \& {Mauro}}{{De Bortoli} et~al.}{2022}]{Bortoli2022}
{De Bortoli} B.~J.,  {Parisi} M.~C.,  {Bassino} L.~P.,  {Geisler} D.,  {Dias} B.,  {Gimeno} G.,  {Angelo} M.~S.,   {Mauro} F.,  2022, \mn@doi [\aap] {10.1051/0004-6361/202243762}, \href {https://ui.adsabs.harvard.edu/abs/2022A&A...664A.168D} {664, A168}

\bibitem[\protect\citeauthoryear{{De Leo}, {Carrera}, {No{\"e}l}, {Read}, {Erkal}  \& {Gallart}}{{De Leo} et~al.}{2020}]{2020deleo}
{De Leo} M.,  {Carrera} R.,  {No{\"e}l} N. E.~D.,  {Read} J.~I.,  {Erkal} D.,   {Gallart} C.,  2020, \mn@doi [\mnras] {10.1093/mnras/staa1122}, \href {https://ui.adsabs.harvard.edu/abs/2020MNRAS.495...98D} {495, 98}

\bibitem[\protect\citeauthoryear{{De Leo}, {Read}, {No{\"e}l}, {Erkal}, {Massana}  \& {Carrera}}{{De Leo} et~al.}{2024}]{2024DeLeo}
{De Leo} M.,  {Read} J.~I.,  {No{\"e}l} N. E.~D.,  {Erkal} D.,  {Massana} P.,   {Carrera} R.,  2024, \mn@doi [\mnras] {10.1093/mnras/stae2428}, \href {https://ui.adsabs.harvard.edu/abs/2024MNRAS.tmp.2351D} {}

\bibitem[\protect\citeauthoryear{{Dias} et~al.,}{{Dias} et~al.}{2021}]{2021Dias}
{Dias} B.,  et~al., 2021, \mn@doi [\aap] {10.1051/0004-6361/202040015}, \href {https://ui.adsabs.harvard.edu/abs/2021A&A...647L...9D} {647, L9}

\bibitem[\protect\citeauthoryear{{Dias} et~al.,}{{Dias} et~al.}{2022}]{2022Dias}
{Dias} B.,  et~al., 2022, \mn@doi [\mnras] {10.1093/mnras/stac259}, \href {https://ui.adsabs.harvard.edu/abs/2022MNRAS.512.4334D} {512, 4334}

\bibitem[\protect\citeauthoryear{{Diaz} \& {Bekki}}{{Diaz} \& {Bekki}}{2011}]{2011PASA...28..117D}
{Diaz} J.,  {Bekki} K.,  2011, \mn@doi [\pasa] {10.1071/AS10044}, \href {https://ui.adsabs.harvard.edu/abs/2011PASA...28..117D} {28, 117}

\bibitem[\protect\citeauthoryear{{Diaz} \& {Bekki}}{{Diaz} \& {Bekki}}{2012}]{2012ApJ...750...36D}
{Diaz} J.~D.,  {Bekki} K.,  2012, \mn@doi [\apj] {10.1088/0004-637X/750/1/36}, \href {https://ui.adsabs.harvard.edu/abs/2012ApJ...750...36D} {750, 36}

\bibitem[\protect\citeauthoryear{{Dobbie}, {Cole}, {Subramaniam}  \& {Keller}}{{Dobbie} et~al.}{2014}]{2014dobbie}
{Dobbie} P.~D.,  {Cole} A.~A.,  {Subramaniam} A.,   {Keller} S.,  2014, \mn@doi [\mnras] {10.1093/mnras/stu926}, \href {https://ui.adsabs.harvard.edu/abs/2014MNRAS.442.1680D} {442, 1680}

\bibitem[\protect\citeauthoryear{{El Youssoufi} et~al.,}{{El Youssoufi} et~al.}{2021}]{2021Youssoufi}
{El Youssoufi} D.,  et~al., 2021, \mn@doi [\mnras] {10.1093/mnras/stab1075}, \href {https://ui.adsabs.harvard.edu/abs/2021MNRAS.505.2020E} {505, 2020}

\bibitem[\protect\citeauthoryear{{Gaia Collaboration} \& et al.}{{Gaia Collaboration} \& et~al.}{2021}]{Gaia2021}
{Gaia Collaboration} et al. 2021, \mn@doi [\aap] {10.1051/0004-6361/202039657}, \href {https://ui.adsabs.harvard.edu/abs/2021A&A...649A...1G} {649, A1}

\bibitem[\protect\citeauthoryear{{Gallart}, {Stetson}, {Meschin}, {Pont}  \& {Hardy}}{{Gallart} et~al.}{2008}]{Gallart2008}
{Gallart} C.,  {Stetson} P.~B.,  {Meschin} I.~P.,  {Pont} F.,   {Hardy} E.,  2008, \mn@doi [\apjl] {10.1086/590552}, \href {https://ui.adsabs.harvard.edu/abs/2008ApJ...682L..89G} {682, L89}

\bibitem[\protect\citeauthoryear{{Glatt} et~al.,}{{Glatt} et~al.}{2008}]{Glatt2008}
{Glatt} K.,  et~al., 2008, \mn@doi [\aj] {10.1088/0004-6256/136/4/1703}, \href {https://ui.adsabs.harvard.edu/abs/2008AJ....136.1703G} {136, 1703}

\bibitem[\protect\citeauthoryear{{Gontcharov} et~al.,}{{Gontcharov} et~al.}{2021}]{2021Gontcharov}
{Gontcharov} G.~A.,  et~al., 2021, \mn@doi [\mnras] {10.1093/mnras/stab2756}, \href {https://ui.adsabs.harvard.edu/abs/2021MNRAS.508.2688G} {508, 2688}

\bibitem[\protect\citeauthoryear{{Gontcharov} et~al.,}{{Gontcharov} et~al.}{2024}]{2024Gontcharov}
{Gontcharov} G.~A.,  et~al., 2024, \mn@doi [arXiv e-prints] {10.48550/arXiv.2404.14797}, \href {https://ui.adsabs.harvard.edu/abs/2024arXiv240414797G} {p. arXiv:2404.14797}

\bibitem[\protect\citeauthoryear{{G{\'o}rski}, {Hivon}, {Banday}, {Wandelt}, {Hansen}, {Reinecke}  \& {Bartelmann}}{{G{\'o}rski} et~al.}{2005}]{2005ApJ...622..759G}
{G{\'o}rski} K.~M.,  {Hivon} E.,  {Banday} A.~J.,  {Wandelt} B.~D.,  {Hansen} F.~K.,  {Reinecke} M.,   {Bartelmann} M.,  2005, \mn@doi [\apj] {10.1086/427976}, \href {http://adsabs.harvard.edu/abs/2005ApJ...622..759G} {622, 759}

\bibitem[\protect\citeauthoryear{{Graczyk} et~al.,}{{Graczyk} et~al.}{2020}]{2020Graczyk}
{Graczyk} D.,  et~al., 2020, \mn@doi [\apj] {10.3847/1538-4357/abbb2b}, \href {https://ui.adsabs.harvard.edu/abs/2020ApJ...904...13G} {904, 13}

\bibitem[\protect\citeauthoryear{{Grady}, {Belokurov}  \& {Evans}}{{Grady} et~al.}{2021}]{2021Grady}
{Grady} J.,  {Belokurov} V.,   {Evans} N.~W.,  2021, \mn@doi [\apj] {10.3847/1538-4357/abd4e4}, \href {https://ui.adsabs.harvard.edu/abs/2021ApJ...909..150G} {909, 150}

\bibitem[\protect\citeauthoryear{{Hatzidimitriou} \& {Hawkins}}{{Hatzidimitriou} \& {Hawkins}}{1989}]{1989MNRAS.241..667H}
{Hatzidimitriou} D.,  {Hawkins} M.~R.~S.,  1989, \mn@doi [\mnras] {10.1093/mnras/241.4.667}, \href {https://ui.adsabs.harvard.edu/abs/1989MNRAS.241..667H} {241, 667}

\bibitem[\protect\citeauthoryear{{Hidalgo} et~al.,}{{Hidalgo} et~al.}{2013}]{Hidalgo2013}
{Hidalgo} S.~L.,  et~al., 2013, \mn@doi [\apj] {10.1088/0004-637X/778/2/103}, \href {https://ui.adsabs.harvard.edu/abs/2013ApJ...778..103H} {778, 103}

\bibitem[\protect\citeauthoryear{{Kalberla} \& {Haud}}{{Kalberla} \& {Haud}}{2015}]{2015Kalberla}
{Kalberla} P.~M.~W.,  {Haud} U.,  2015, \mn@doi [\aap] {10.1051/0004-6361/201525859}, \href {https://ui.adsabs.harvard.edu/abs/2015A&A...578A..78K} {578, A78}

\bibitem[\protect\citeauthoryear{{Kirby}, {Cohen}, {Guhathakurta}, {Cheng}, {Bullock}  \& {Gallazzi}}{{Kirby} et~al.}{2013}]{2013Kirby}
{Kirby} E.~N.,  {Cohen} J.~G.,  {Guhathakurta} P.,  {Cheng} L.,  {Bullock} J.~S.,   {Gallazzi} A.,  2013, \mn@doi [\apj] {10.1088/0004-637X/779/2/102}, \href {https://ui.adsabs.harvard.edu/abs/2013ApJ...779..102K} {779, 102}

\bibitem[\protect\citeauthoryear{{Kirby}, {Rizzi}, {Held}, {Cohen}, {Cole}, {Manning}, {Skillman}  \& {Weisz}}{{Kirby} et~al.}{2017}]{2017Kirby}
{Kirby} E.~N.,  {Rizzi} L.,  {Held} E.~V.,  {Cohen} J.~G.,  {Cole} A.~A.,  {Manning} E.~M.,  {Skillman} E.~D.,   {Weisz} D.~R.,  2017, \mn@doi [\apj] {10.3847/1538-4357/834/1/9}, \href {https://ui.adsabs.harvard.edu/abs/2017ApJ...834....9K} {834, 9}

\bibitem[\protect\citeauthoryear{{Leaman}, {VandenBerg}  \& {Mendel}}{{Leaman} et~al.}{2013}]{2013Leaman}
{Leaman} R.,  {VandenBerg} D.~A.,   {Mendel} J.~T.,  2013, \mn@doi [\mnras] {10.1093/mnras/stt1540}, \href {https://ui.adsabs.harvard.edu/abs/2013MNRAS.436..122L} {436, 122}

\bibitem[\protect\citeauthoryear{{Li}, {Jiang}  \& {Ren}}{{Li} et~al.}{2024}]{LI2024}
{Li} Y.,  {Jiang} B.,   {Ren} Y.,  2024, \mn@doi [\aj] {10.3847/1538-3881/ad23e8}, \href {https://ui.adsabs.harvard.edu/abs/2024AJ....167..123L} {167, 123}

\bibitem[\protect\citeauthoryear{{Majewski} \& et al.}{{Majewski} \& et~al.}{2017}]{2017Majewski}
{Majewski} S.~R.,  et al. 2017, \mn@doi [\aj] {10.3847/1538-3881/aa784d}, \href {https://ui.adsabs.harvard.edu/abs/2017AJ....154...94M} {154, 94}

\bibitem[\protect\citeauthoryear{{Mart{\'\i}nez-V{\'a}zquez}, {Salinas}  \& {Vivas}}{{Mart{\'\i}nez-V{\'a}zquez} et~al.}{2021}]{2021Martinez}
{Mart{\'\i}nez-V{\'a}zquez} C.~E.,  {Salinas} R.,   {Vivas} A.~K.,  2021, \mn@doi [\aj] {10.3847/1538-3881/abd55e}, \href {https://ui.adsabs.harvard.edu/abs/2021AJ....161..120M} {161, 120}

\bibitem[\protect\citeauthoryear{{Massana} et~al.,}{{Massana} et~al.}{2020}]{2020Massana}
{Massana} P.,  et~al., 2020, \mn@doi [\mnras] {10.1093/mnras/staa2451}, \href {https://ui.adsabs.harvard.edu/abs/2020MNRAS.498.1034M} {498, 1034}

\bibitem[\protect\citeauthoryear{{McQuinn} et~al.,}{{McQuinn} et~al.}{2017}]{McQuinn2017}
{McQuinn} K. B.~W.,  et~al., 2017, \mn@doi [\apj] {10.3847/1538-4357/834/1/78}, \href {https://ui.adsabs.harvard.edu/abs/2017ApJ...834...78M} {834, 78}

\bibitem[\protect\citeauthoryear{{Mucciarelli}, {Minelli}, {Bellazzini}, {Lardo}, {Romano}, {Origlia}  \& {Ferraro}}{{Mucciarelli} et~al.}{2023a}]{mucciarelli2023}
{Mucciarelli} A.,  {Minelli} A.,  {Bellazzini} M.,  {Lardo} C.,  {Romano} D.,  {Origlia} L.,   {Ferraro} F.~R.,  2023a, \mn@doi [\aap] {10.1051/0004-6361/202245133}, \href {https://ui.adsabs.harvard.edu/abs/2023A&A...671A.124M} {671, A124}

\bibitem[\protect\citeauthoryear{{Mucciarelli}, {Minelli}, {Lardo}, {Massari}, {Bellazzini}, {Romano}, {Origlia}  \& {Ferraro}}{{Mucciarelli} et~al.}{2023b}]{mucciarelli2023c}
{Mucciarelli} A.,  {Minelli} A.,  {Lardo} C.,  {Massari} D.,  {Bellazzini} M.,  {Romano} D.,  {Origlia} L.,   {Ferraro} F.~R.,  2023b, \mn@doi [\aap] {10.1051/0004-6361/202347120}, \href {https://ui.adsabs.harvard.edu/abs/2023A&A...677A..61M} {677, A61}

\bibitem[\protect\citeauthoryear{{Mucciarelli}, {Minelli}, {Lardo}, {Massari}, {Bellazzini}, {Romano}, {Origlia}  \& {Ferraro}}{{Mucciarelli} et~al.}{2023c}]{2023Mucciarelli}
{Mucciarelli} A.,  {Minelli} A.,  {Lardo} C.,  {Massari} D.,  {Bellazzini} M.,  {Romano} D.,  {Origlia} L.,   {Ferraro} F.~R.,  2023c, \mn@doi [\aap] {10.1051/0004-6361/202347120}, \href {https://ui.adsabs.harvard.edu/abs/2023A&A...677A..61M} {677, A61}

\bibitem[\protect\citeauthoryear{{Myers} et~al.,}{{Myers} et~al.}{2022}]{2022Myers}
{Myers} N.,  et~al., 2022, \mn@doi [\aj] {10.3847/1538-3881/ac7ce5}, \href {https://ui.adsabs.harvard.edu/abs/2022AJ....164...85M} {164, 85}

\bibitem[\protect\citeauthoryear{Newville, Stensitzki, Allen  \& Ingargiola}{Newville et~al.}{2014}]{lmfit}
Newville M.,  Stensitzki T.,  Allen D.~B.,   Ingargiola A.,  2014, \mn@doi [Journal of Open Source Software] {10.21105/joss.00774}, 3, 774

\bibitem[\protect\citeauthoryear{{Nidever}, {Majewski}  \& {Butler Burton}}{{Nidever} et~al.}{2008}]{2008Nidever}
{Nidever} D.~L.,  {Majewski} S.~R.,   {Butler Burton} W.,  2008, \mn@doi [\apj] {10.1086/587042}, \href {https://ui.adsabs.harvard.edu/abs/2008ApJ...679..432N} {679, 432}

\bibitem[\protect\citeauthoryear{{Nidever}, {Majewski}, {Butler Burton}  \& {Nigra}}{{Nidever} et~al.}{2010}]{2010Nidever}
{Nidever} D.~L.,  {Majewski} S.~R.,  {Butler Burton} W.,   {Nigra} L.,  2010, \mn@doi [\apj] {10.1088/0004-637X/723/2/1618}, \href {https://ui.adsabs.harvard.edu/abs/2010ApJ...723.1618N} {723, 1618}

\bibitem[\protect\citeauthoryear{{Nidever}, {Monachesi}, {Bell}, {Majewski}, {Mu{\~n}oz}  \& {Beaton}}{{Nidever} et~al.}{2013}]{2013Nidever}
{Nidever} D.~L.,  {Monachesi} A.,  {Bell} E.~F.,  {Majewski} S.~R.,  {Mu{\~n}oz} R.~R.,   {Beaton} R.~L.,  2013, \mn@doi [\apj] {10.1088/0004-637X/779/2/145}, \href {https://ui.adsabs.harvard.edu/abs/2013ApJ...779..145N} {779, 145}

\bibitem[\protect\citeauthoryear{{Nidever}, {Hasselquist}, {Hayes}  \& et al.}{{Nidever} et~al.}{2020}]{nidever2020}
{Nidever} D.~L.,  {Hasselquist} S.,  {Hayes} C.~R.,   et al. 2020, \mn@doi [\apj] {10.3847/1538-4357/ab7305}, \href {https://ui.adsabs.harvard.edu/abs/2020ApJ...895...88N} {895, 88}

\bibitem[\protect\citeauthoryear{{No{\"e}l}, {Aparicio}, {Gallart}, {Hidalgo}, {Costa}  \& {M{\'e}ndez}}{{No{\"e}l} et~al.}{2009}]{Noel2009}
{No{\"e}l} N. E.~D.,  {Aparicio} A.,  {Gallart} C.,  {Hidalgo} S.~L.,  {Costa} E.,   {M{\'e}ndez} R.~A.,  2009, \mn@doi [\apj] {10.1088/0004-637X/705/2/1260}, \href {https://ui.adsabs.harvard.edu/abs/2009ApJ...705.1260N} {705, 1260}

\bibitem[\protect\citeauthoryear{{No{\"e}l}, {Greggio}, {Renzini}, {Carollo}  \& {Maraston}}{{No{\"e}l} et~al.}{2013}]{Noel2013b}
{No{\"e}l} N.~E.~D.,  {Greggio} L.,  {Renzini} A.,  {Carollo} C.~M.,   {Maraston} C.,  2013, \mn@doi [\apj] {10.1088/0004-637X/772/1/58}, \href {https://ui.adsabs.harvard.edu/abs/2013ApJ...772...58N} {772, 58}

\bibitem[\protect\citeauthoryear{{No{\"e}l}, {Conn}, {Read}, {Carrera}, {Dolphin}  \& {Rix}}{{No{\"e}l} et~al.}{2015}]{Noel2015}
{No{\"e}l} N.~E.~D.,  {Conn} B.~C.,  {Read} J.~I.,  {Carrera} R.,  {Dolphin} A.,   {Rix} H.~W.,  2015, \mn@doi [\mnras] {10.1093/mnras/stv1614}, \href {https://ui.adsabs.harvard.edu/abs/2015MNRAS.452.4222N} {452, 4222}

\bibitem[\protect\citeauthoryear{{Olsen}, {Zaritsky}, {Blum}, {Boyer}  \& {Gordon}}{{Olsen} et~al.}{2011}]{Olsen2011}
{Olsen} K. A.~G.,  {Zaritsky} D.,  {Blum} R.~D.,  {Boyer} M.~L.,   {Gordon} K.~D.,  2011, \mn@doi [\apj] {10.1088/0004-637X/737/1/29}, \href {https://ui.adsabs.harvard.edu/abs/2011ApJ...737...29O} {737, 29}

\bibitem[\protect\citeauthoryear{{Parisi}, {Grocholski}, {Geisler}, {Sarajedini}  \& {Clari{\'a}}}{{Parisi} et~al.}{2009}]{Parisi2009}
{Parisi} M.~C.,  {Grocholski} A.~J.,  {Geisler} D.,  {Sarajedini} A.,   {Clari{\'a}} J.~J.,  2009, \mn@doi [\aj] {10.1088/0004-6256/138/2/517}, \href {https://ui.adsabs.harvard.edu/abs/2009AJ....138..517P} {138, 517}

\bibitem[\protect\citeauthoryear{{Parisi}, {Geisler}, {Grocholski}, {Clari{\'a}}  \& {Sarajedini}}{{Parisi} et~al.}{2010}]{Parisi2010}
{Parisi} M.~C.,  {Geisler} D.,  {Grocholski} A.~J.,  {Clari{\'a}} J.~J.,   {Sarajedini} A.,  2010, \mn@doi [\aj] {10.1088/0004-6256/139/3/1168}, \href {https://ui.adsabs.harvard.edu/abs/2010AJ....139.1168P} {139, 1168}

\bibitem[\protect\citeauthoryear{{Parisi}, {Geisler}, {Clari{\'a}}, {Villanova}, {Marcionni}, {Sarajedini}  \& {Grocholski}}{{Parisi} et~al.}{2015}]{Parisi2015}
{Parisi} M.~C.,  {Geisler} D.,  {Clari{\'a}} J.~J.,  {Villanova} S.,  {Marcionni} N.,  {Sarajedini} A.,   {Grocholski} A.~J.,  2015, \mn@doi [\aj] {10.1088/0004-6256/149/5/154}, \href {https://ui.adsabs.harvard.edu/abs/2015AJ....149..154P} {149, 154}

\bibitem[\protect\citeauthoryear{{Parisi}, {Geisler}, {Carraro}, {Clari{\'a}}, {Villanova}, {Gramajo}, {Sarajedini}  \& {Grocholski}}{{Parisi} et~al.}{2016}]{Parisi2016}
{Parisi} M.~C.,  {Geisler} D.,  {Carraro} G.,  {Clari{\'a}} J.~J.,  {Villanova} S.,  {Gramajo} L.~V.,  {Sarajedini} A.,   {Grocholski} A.~J.,  2016, \mn@doi [\aj] {10.3847/0004-6256/152/3/58}, \href {https://ui.adsabs.harvard.edu/abs/2016AJ....152...58P} {152, 58}

\bibitem[\protect\citeauthoryear{{Parisi}, {Gramajo}, {Geisler}, {Dias}, {Clari{\'a}}, {Da Costa}  \& {Grebel}}{{Parisi} et~al.}{2022}]{Parisi2022}
{Parisi} M.~C.,  {Gramajo} L.~V.,  {Geisler} D.,  {Dias} B.,  {Clari{\'a}} J.~J.,  {Da Costa} G.,   {Grebel} E.~K.,  2022, \mn@doi [\aap] {10.1051/0004-6361/202142597}, \href {https://ui.adsabs.harvard.edu/abs/2022A&A...662A..75P} {662, A75}

\bibitem[\protect\citeauthoryear{{Patel}, {Besla}  \& {Mandel}}{{Patel} et~al.}{2017}]{2017MNRAS.468.3428P}
{Patel} E.,  {Besla} G.,   {Mandel} K.,  2017, \mn@doi [\mnras] {10.1093/mnras/stx698}, \href {https://ui.adsabs.harvard.edu/abs/2017MNRAS.468.3428P} {468, 3428}

\bibitem[\protect\citeauthoryear{{Piatti}, {Sarajedini}, {Geisler}, {Clark}  \& {Seguel}}{{Piatti} et~al.}{2007}]{Piatti2007}
{Piatti} A.~E.,  {Sarajedini} A.,  {Geisler} D.,  {Clark} D.,   {Seguel} J.,  2007, \mn@doi [\mnras] {10.1111/j.1365-2966.2007.11604.x}, \href {https://ui.adsabs.harvard.edu/abs/2007MNRAS.377..300P} {377, 300}

\bibitem[\protect\citeauthoryear{{Planck Collaboration} \& et al.}{{Planck Collaboration} \& et~al.}{2016}]{Planck2016}
{Planck Collaboration} et al. 2016, \mn@doi [\aap] {10.1051/0004-6361/201629022}, \href {https://ui.adsabs.harvard.edu/abs/2016A&A...596A.109P} {596, A109}

\bibitem[\protect\citeauthoryear{{Putman}, {Staveley-Smith}, {Freeman}, {Gibson}  \& {Barnes}}{{Putman} et~al.}{2003}]{2003ApJ...586..170P}
{Putman} M.~E.,  {Staveley-Smith} L.,  {Freeman} K.~C.,  {Gibson} B.~K.,   {Barnes} D.~G.,  2003, \mn@doi [\apj] {10.1086/344477}, \href {https://ui.adsabs.harvard.edu/abs/2003ApJ...586..170P} {586, 170}

\bibitem[\protect\citeauthoryear{{Radburn-Smith} et~al.,}{{Radburn-Smith} et~al.}{2012}]{2012Radburn}
{Radburn-Smith} D.~J.,  et~al., 2012, \mn@doi [\apj] {10.1088/0004-637X/753/2/138}, \href {https://ui.adsabs.harvard.edu/abs/2012ApJ...753..138R} {753, 138}

\bibitem[\protect\citeauthoryear{{Randich}, {Gilmore}, {Magrini}  et~al.}{{Randich} et~al.}{2022}]{2022A&A...666A.121R}
{Randich} S.,  {Gilmore} G.,  {Magrini} L.,   et~al., 2022, \mn@doi [\aap] {10.1051/0004-6361/202243141}, \href {https://ui.adsabs.harvard.edu/abs/2022A&A...666A.121R} {666, A121}

\bibitem[\protect\citeauthoryear{{Ro{\v{s}}kar}, {Debattista}, {Quinn}, {Stinson}  \& {Wadsley}}{{Ro{\v{s}}kar} et~al.}{2008}]{2008Ro}
{Ro{\v{s}}kar} R.,  {Debattista} V.~P.,  {Quinn} T.~R.,  {Stinson} G.~S.,   {Wadsley} J.,  2008, \mn@doi [\apjl] {10.1086/592231}, \href {https://ui.adsabs.harvard.edu/abs/2008ApJ...684L..79R} {684, L79}

\bibitem[\protect\citeauthoryear{{Ro{\v{s}}kar}, {Debattista}, {Quinn}  \& {Wadsley}}{{Ro{\v{s}}kar} et~al.}{2012}]{2012Ro}
{Ro{\v{s}}kar} R.,  {Debattista} V.~P.,  {Quinn} T.~R.,   {Wadsley} J.,  2012, \mn@doi [\mnras] {10.1111/j.1365-2966.2012.21860.x}, \href {https://ui.adsabs.harvard.edu/abs/2012MNRAS.426.2089R} {426, 2089}

\bibitem[\protect\citeauthoryear{{Rutledge}, {Hesser}, {Stetson}, {Mateo}, {Simard}, {Bolte}, {Friel}  \& {Copin}}{{Rutledge} et~al.}{1997a}]{1997Rutledgeb}
{Rutledge} G.~A.,  {Hesser} J.~E.,  {Stetson} P.~B.,  {Mateo} M.,  {Simard} L.,  {Bolte} M.,  {Friel} E.~D.,   {Copin} Y.,  1997a, \mn@doi [\pasp] {10.1086/133958}, \href {https://ui.adsabs.harvard.edu/abs/1997PASP..109..883R} {109, 883}

\bibitem[\protect\citeauthoryear{{Rutledge}, {Hesser}  \& {Stetson}}{{Rutledge} et~al.}{1997b}]{1997Rutledgea}
{Rutledge} G.~A.,  {Hesser} J.~E.,   {Stetson} P.~B.,  1997b, \mn@doi [\pasp] {10.1086/133959}, \href {https://ui.adsabs.harvard.edu/abs/1997PASP..109..907R} {109, 907}

\bibitem[\protect\citeauthoryear{{Sakowska} et~al.,}{{Sakowska} et~al.}{2024}]{2024Sakowska}
{Sakowska} J.~D.,  et~al., 2024, \mn@doi [\mnras] {10.1093/mnras/stae1766}, \href {https://ui.adsabs.harvard.edu/abs/2024MNRAS.532.4272S} {532, 4272}

\bibitem[\protect\citeauthoryear{{Saunders}, {Bridges}, {Gillingham}  et~al.}{{Saunders} et~al.}{2004}]{2004SPIE.5492..389S}
{Saunders} W.,  {Bridges} T.,  {Gillingham} P.,   et~al., 2004, in {Moorwood} A. F.~M.,  {Iye} M.,  eds,  Society of Photo-Optical Instrumentation Engineers (SPIE) Conference Series Vol. 5492, Ground-based Instrumentation for Astronomy. pp 389--400, \mn@doi{10.1117/12.550871}

\bibitem[\protect\citeauthoryear{{Sharp} \& {Birchall}}{{Sharp} \& {Birchall}}{2010}]{Sharp2010}
{Sharp} R.,  {Birchall} M.~N.,  2010, \mn@doi [\pasa] {10.1071/AS08001}, \href {https://ui.adsabs.harvard.edu/abs/2010PASA...27...91S} {27, 91}

\bibitem[\protect\citeauthoryear{{Skrutskie} et~al.,}{{Skrutskie} et~al.}{2006}]{2006AJ....131.1163S}
{Skrutskie} M.~F.,  et~al., 2006, \mn@doi [\aj] {10.1086/498708}, \href {https://ui.adsabs.harvard.edu/abs/2006AJ....131.1163S} {131, 1163}

\bibitem[\protect\citeauthoryear{{Smitka} \& {Layden}}{{Smitka} \& {Layden}}{2010}]{2010Smitka}
{Smitka} M.~T.,  {Layden} A.~C.,  2010, in American Astronomical Society Meeting Abstracts \#215. p. 606.18

\bibitem[\protect\citeauthoryear{{Starkenburg} et~al.,}{{Starkenburg} et~al.}{2010}]{Starkenburg2010}
{Starkenburg} E.,  et~al., 2010, \mn@doi [\aap] {10.1051/0004-6361/200913759}, \href {https://ui.adsabs.harvard.edu/abs/2010A&A...513A..34S} {513, A34}

\bibitem[\protect\citeauthoryear{{Subramanian} \& {Subramaniam}}{{Subramanian} \& {Subramaniam}}{2012}]{2012Subramanian}
{Subramanian} S.,  {Subramaniam} A.,  2012, \mn@doi [\apj] {10.1088/0004-637X/744/2/128}, \href {https://ui.adsabs.harvard.edu/abs/2012ApJ...744..128S} {744, 128}

\bibitem[\protect\citeauthoryear{{Taibi}, {Battaglia}, {Leaman}, {Brooks}, {Riggs}, {Munshi}, {Revaz}  \& {Jablonka}}{{Taibi} et~al.}{2022}]{2022Taibi}
{Taibi} S.,  {Battaglia} G.,  {Leaman} R.,  {Brooks} A.,  {Riggs} C.,  {Munshi} F.,  {Revaz} Y.,   {Jablonka} P.,  2022, \mn@doi [\aap] {10.1051/0004-6361/202243508}, \href {https://ui.adsabs.harvard.edu/abs/2022A&A...665A..92T} {665, A92}

\bibitem[\protect\citeauthoryear{{Tatton} et~al.,}{{Tatton} et~al.}{2021}]{2021Tatton}
{Tatton} B.~L.,  et~al., 2021, \mn@doi [\mnras] {10.1093/mnras/staa3857}, \href {https://ui.adsabs.harvard.edu/abs/2021MNRAS.504.2983T} {504, 2983}

\bibitem[\protect\citeauthoryear{{Van der Swaelmen}, {Hill}, {Primas}  \& {Cole}}{{Van der Swaelmen} et~al.}{2013}]{vanderswaelmen2013}
{Van der Swaelmen} M.,  {Hill} V.,  {Primas} F.,   {Cole} A.~A.,  2013, \mn@doi [\aap] {10.1051/0004-6361/201321109}, \href {https://ui.adsabs.harvard.edu/abs/2013A&A...560A..44V} {560, A44}

\bibitem[\protect\citeauthoryear{{Vargas}, {Villanova}, {Geisler}, {Mu{\~n}oz}, {Monaco}, {O'Connell}  \& {Sarajedini}}{{Vargas} et~al.}{2022}]{2022Vargas}
{Vargas} C.,  {Villanova} S.,  {Geisler} D.,  {Mu{\~n}oz} C.,  {Monaco} L.,  {O'Connell} J.,   {Sarajedini} A.,  2022, \mn@doi [\mnras] {10.1093/mnras/stac1758}, \href {https://ui.adsabs.harvard.edu/abs/2022MNRAS.515.1903V} {515, 1903}

\bibitem[\protect\citeauthoryear{{Williams}, {Dalcanton}, {Dolphin}, {Holtzman}  \& {Sarajedini}}{{Williams} et~al.}{2009}]{2009Williams}
{Williams} B.~F.,  {Dalcanton} J.~J.,  {Dolphin} A.~E.,  {Holtzman} J.,   {Sarajedini} A.,  2009, \mn@doi [\apjl] {10.1088/0004-637X/695/1/L15}, \href {https://ui.adsabs.harvard.edu/abs/2009ApJ...695L..15W} {695, L15}

\bibitem[\protect\citeauthoryear{Zonca, Singer, Lenz, Reinecke, Rosset, Hivon  \& Gorski}{Zonca et~al.}{2019}]{Zonca2019}
Zonca A.,  Singer L.,  Lenz D.,  Reinecke M.,  Rosset C.,  Hivon E.,   Gorski K.,  2019, \mn@doi [Journal of Open Source Software] {10.21105/joss.01298}, 4, 1298

\bibitem[\protect\citeauthoryear{{van der Marel}, {Alves}, {Hardy}  \& {Suntzeff}}{{van der Marel} et~al.}{2002}]{2002AJ....124.2639V}
{van der Marel} R.~P.,  {Alves} D.~R.,  {Hardy} E.,   {Suntzeff} N.~B.,  2002, \mn@doi [\aj] {10.1086/343775}, \href {https://ui.adsabs.harvard.edu/abs/2002AJ....124.2639V} {124, 2639}

\makeatother
\end{thebibliography}
